\documentclass[AMA,STIX1COL]{WileyNJD-v2}

\articletype{RESEARCH ARTICLE}%

\received{Day April 2016}
\revised{6 June 2016}
\accepted{6 June 2016}

\begin{document}

\title{Hypothesis tests for multiple responses regression: effect of probiotics on addiction and binge eating disorder.}

\author[1]{Lineu Alberto Cavazani de Freitas}

\author[2]{Ligia de Oliveira Carlos}

\author[2]{Antônio Carlos Ligocki Campos}

\author[3]{Wagner Hugo Bonat}

\authormark{FREITAS \textsc{et al}}

\address[1]{\orgdiv{Department of Informatics}, \orgname{Paraná Federal University}, \orgaddress{\state{Paraná}, \country{Brazil}}}

\address[2]{\orgdiv{Department of Surgical Clinic}, \orgname{Paraná Federal University}, \orgaddress{\state{Paraná}, \country{Brazil}}}

\address[3]{\orgdiv{Department of Statistics}, \orgname{Paraná Federal University}, \orgaddress{\state{Paraná}, \country{Brazil}}}

\corres{Lineu Alberto Cavazani de Freitas, Department of Informatics, Paraná Federal University, R. Evaristo F. Ferreira da Costa, 383-391, Jardim das Américas, Curitiba, Paraná, 82590-300, Brazil. \email{lineuacf@gmail.com}}

\abstract[Summary]{

Clinical trials are common in medical research where multiple non-Gaussian responses and time-dependent observations are frequent.
The analysis of data from these studies requires statistical modeling techniques that take these characteristics into account. 
We propose a general strategy based on the Wald statistics to perform hypothesis tests like ANOVAs, MANOVAs and multiple comparison tests on 
regression and dispersion parameters of multivariate covariance generalized linear models (McGLMs). 
McGLMs provide a general statistical modeling framework for normal and non-normal multivariate data analysis along with a wide range of correlation structures. 
We design different simulation scenarios to verify the properties of the proposed tests. 
The results are promising showing that the proposed tests present the levels of confidence close to the specified one for all simulation study scenarios.
Complementary to the proposal, we developed implementations in the R language to carry out the tests presented, the codes are available in the supplementary material.
The proposal is motivated by the analysis of a clinical trial that aims to evaluate the effect of the use of probiotics in the control of addiction and binge eating disorder in patients undergoing bariatric surgery. 
The subjects were separated into two groups (placebo and treatment) and evaluated at three different times. 
The results indicate that addiction and binge eating disorder reduce over time, but there is no difference between groups at each time point.

}



\maketitle



\section{Introduction}\label{sec1}


Clinical trials represent one of the most common procedures used to assess whether a new treatment is safe, effective or even better than a standard treatment. These are procedures widely used in medical research with the aim of evaluating interventions on a group of individuals. In general, the data generated by clinical trials are not trivial and, for analysis, there is a need for statistical methods suitable to deal with all the specifics of this type of study.

According to Meinert and Tonascia \cite{meinert1986clinical} "a clinical trial is a planned experiment designed to assess the efficacy of a treatment in man by comparing the outcomes in a group of patients treated with the test with those observed in a comparable group of patients receiving a control treatment where patients in both groups are enrolled, treated, and followed over the same time period".

In this type of study participants are randomly assigned to a treatment or control group (or to multiple treatment groups) \cite{hannan2008randomized} and the objective is to assess whether there is a significant difference between the groups.


Often, in medical studies that use clinical trials, multiple outcomes are taken for a group of patients, for example: Kangovi et al \cite{kangovi2018effect} evaluated the effect of community health worker support on clinical outcomes of low-income patients; Song and Baicker \cite{song2019effect} studied the effect of a workplace wellness program on employee health and economic outcomes; Thyregod et al \cite{thyregod2019five} presents the results of a five-year clinical and echocardiographic outcomes from the nordic aortic valve intervention (NOTION); Schmitz et al \cite{schmitz2019effect} evaluated the effect of Home-Based Exercise and Weight Loss Programs on Breast Cancer–Related Lymphedema Outcomes; Bonat et al \cite{plastica} proposed a model to deal with multiple outcomes in repeated measures studies motivated by a problem that aimed to compare different aesthetic eyelid surgery techniques; Petterle et al \cite{petterle2021multivariate} proposed a multivariate regression model to handle multiple continuous bounded outcomes motivated by a body fat percentage data. Consequently, the joint analysis of multiple outcomes studies has been of increased interest in the medical and statistical literature. 


Similarly to the aforementioned articles, we are interested in the analysis of multiple outcomes in the context of longitudinal data analysis. The study we shall describe was conducted with the objective of evaluating the effect of the use of probiotics in the control of addiction and binge eating disorder in patients undergoing bariatric surgery. This is a clinical trial with two response variables: a score that characterizes compulsion and the number of symptoms presented that characterize addiction. Both responses are limited counts and for analysis purposes have been transformed to unit range.

In this study, a set of subjects was divided into two groups: one of them received a placebo and the other received the treatment. In addition, individuals were evaluated at 3 different time points: the first evaluation was performed approximately 10 days before surgery; follow-up assessments were performed approximately three months and one year postoperatively.


It is important to note that this is a multivariate problem (contains two response variables), with non-Gaussian responses (both taking values in the unit range) and that the observations that make up the data set cannot be considered independent, considering that measurements taken on the same individual at different time points tend to be correlated. Therefore, it is a problem that standard modeling techniques would be difficult to apply. It is necessary to use some methodology that supports the design of the clinical trial adequately.

In this article, we adopt the multivariate covariance generalized linear models (McGLM) framework \cite{Bonat16}, which provides an environment for modeling multiple non-Gaussian responses simultaneously with flexible and interpretable modeling of the covariance structure. The within outcomes covariance matrix is specified for each marginal outcome using a linear combination of known matrices, while the joint covariance matrix is specified using the generalized Kronecker product \cite{martinez13,Bonat16}.


When working with regression models, a common interest is to verify if the absence of a certain explanatory variable of the model would generate a loss in the adjustment. Thus, a conjecture of interest is to assess whether there is sufficient evidence in the data to state that a given explanatory variable has no effect on the response. This is done through hypothesis testing. In this context, three hypothesis tests are common: the likelihood ratio test, the Wald test, and the Lagrange multiplier test, also known as the score test. Engle \cite{engle} describes the general formulation of the three tests. All of them are based on the likelihood function of the models. 


In the case of standard linear models, there are techniques such as analysis of variance (ANOVA), initially proposed by Fisher and Mackenzie \cite{anova_fisher}. According to St, Wold et al \cite{anova1}, ANOVA is one of the most widely used statistical methods to test hypotheses and it is present in virtually all introductory statistics materials. The objective of the technique is to evaluate the effect of each explanatory variables on the response. This is done by comparing models with and without each of the explanatory variables. Therefore, this procedure makes it possible to assess whether the removal of each of the variables generates a significantly worse model when compared to the model with the variable. For the multivariate case, the technique of analysis of variance (ANOVA) is extended to the multivariate analysis of variance \citep{manova}, MANOVA. Among the multivariate hypothesis tests already discussed in the literature for MANOVAs, we have the Wilk's \cite{wilks} $\lambda$, Hotelling-Lawley trace \cite{lawley,hotelling}, Pillai trace \cite{pillai} and largest root of Roy \cite{roy}. 


Complementary to ANOVAs and MANOVAs are multiple comparison tests. Such procedures are used when the analysis of variance points to the existence of a significant effect of the parameters associated with a categorical variable, that is, there is at least one significant difference between the levels of a factor. Thus, the multiple comparison test is another procedure based on hypothesis testing, used to determine where these differences appear. For example, suppose there is a three-level categorical variable $X$ in the model that assumes the levels: A, B, or C. The analysis of variance will show if there is an effect of the variable $X$ in the model, that is, if the response values are associated with the levels of $X$, however this result will not tell us whether the response values differ from A to B, or from A to C, or whether B differs from C. To detect such differences, multiple comparison tests are used, some of the tests discussed in the literature are: the Dunnett, Tukey, student's t test (LSD), Scott-Knott test, among others. Hsu \cite{hsu1996multiple} discusses various procedures for the purpose of multiple comparisons. Bretz, Hothorn, and Westfall \cite{bretz2008multiple} presents procedures for multiple comparisons in linear models.


However, for multiple response regression models, there are few alternatives for hypothesis testing. Therefore, as it is a flexible class of models with high power of application to practical problems, our general objective is the development of hypothesis tests for McGLMs. We have the following specific objectives: to propose the use of the Wald test to carry out tests of general hypotheses on regression and dispersion parameters of McGLMs, also enabling the generation of tables of analysis of variance, multivariate analysis of variance and multiple comparisons tests for regression models with multiple Gaussian and non-Gaussian responses. Another objective is to evaluate the properties of the proposed tests based on simulation studies and to motivate the potential application of the discussed methodologies based on the analysis of the aforementioned dataset. In the case study, standard modeling techniques would be difficult to apply, however it is a problem of possible analysis via McGLM and hypothesis tests can be used to evaluate the effect of the interaction between time and use of the probiotic on addiction and binge eating disorder.


We present the dataset in \autoref{sec2}. In \autoref{sec3} we present the review of the general structure and estimation of the parameters of a McGLM, based on the ideas of Bonat and Jørgensen \cite{Bonat16}. In \autoref{sec4} our proposal is presented with the details of the Wald test to evaluate assumptions about parameters of a McGLM. In \autoref{sec5} we present the results of the performance evaluation of the proposed test based on a simulation study. In the \autoref{sec6} we apply the model and present the main results of the analysis of the data from the clinical study that aims to evaluate the use of probiotics in the control of addiction and binge eating disorder in patients undergoing bariatric surgery. Finally, the main results are discussed in \autoref{sec7}, including some directions for future investigations. The data sets and R codes are available online and can be assessed in \href{https://github.com/lineu96/paper-htmcglm}{https://github.com/lineu96/paper-htmcglm}.


\section{Data set}\label{sec2}


This is a randomized, double-blind, placebo-controlled clinical trial conducted with patients undergoing Roux-en-Y Gastric Bypass (RYGB) from April 2018 to December 2019. The study was approved by the Research Ethics Committee of the Pontifical Catholic University of Paraná (PUCPR) (nº 4.252.808 ) and registered by the Brazilian Registry of Clinical Trials - REBEC (nºRBR-4x3gqp). The research was explained to each participant before their participation and, from those who agreed, written informed consent was obtained. The division of groups (placebo or probiotic) was done randomly. The inclusion criteria for subjects in the study were: adults (18-59 years) who would undergo RYGB, with a body mass index (BMI) $\geq$ 35 kg/m2 and who did not use antibiotics before surgery. 

Patients who underwent other surgical techniques or reoperation, had post-surgical complications, received antibiotic therapy concomitantly with the use of probiotic/placebo or did not use probiotic/placebo tablets for more than nine consecutive days were excluded from the study. Both groups received the same dietary guidelines after surgery, were followed up by the same surgical team (doctor, nutritionist and psychologist) and had the same number of pre-scheduled consultations before and after surgery, following the protocol established by the institution where the study was carried out.

On the seventh postoperative day, participants were instructed to ingest two chewable tablets/day of placebo or Flora Vantage probiotic tablet, 5 billion of Lactobacillus acidophilus NCFM \textregistered Strain and 5 billion of Bifidobacterium lactis Bi-07 \textregistered) from Bariatric Advantage (Aliso Viejo, CA, USA) for 90 days. Subjects were evaluated at 3 time points. The first evaluation (T0) was performed approximately 10 days before surgery. Follow-up assessments were performed approximately three months (T1) and one year postoperatively (T2). At these moments, clinical and anthropometric assessments were performed, as well as self-administered questionnaires were given to the participants at each meeting.

The evaluation of binge eating was based on the binge eating scale (BES), one of the most used tools to measure binge eating in research, with numerous studies proving its effectiveness, translated into Portuguese and validated for individuals with obesity and submitted to bariatric surgery. This is a 16-item Likert scale questionnaire, prepared according to the Diagnostic and Statistical Manual of Mental Disorders (3rd edition) \cite{spitzer1980diagnostic} by Gormally et al \cite{gormally1982assessment}. The individuals were instructed to select the option that best represented their answer and the final score was the sum of the points for each item, this score ranging from 0 to 46.

To assess food addiction, the Food Addiction Scale (YFAS) was used, a questionnaire that seeks to detect symptoms of addictive eating behaviors. YFAS was based on substance dependence criteria from Diagnostic and Statistical Manual IV – Proofreading (DSM-IV-TR) \cite{segal2010diagnostic} and endorsed for highly processed foods. This questionnaire was developed by Gearhardt et al \cite{gearhardt2009preliminary} and is a combination of 25 Likert scale options and the evaluation option used was the number of addiction symptoms.


The final sample consists of 71 individuals: 33 belong to the placebo group and 38 to the treatment group. If all these individuals were evaluated at the 3 time points defined in the study, the dataset would have 213 observations. However, throughout the study, several individuals did not attend the consultations, which causes missing data in the dataset. After processing the data and excluding missing observations, 184 observations remained.


For analysis purposes, the score that characterizes compulsion and the number of symptoms presented that characterize addiction were transformed to the unitary scale, considering that these are counts variables restricted to a range. The purpose of the analysis is to assess the effect of moment and group on addiction and compulsion metrics. The dataset contains the following variables:

\begin{itemize}
  \item id: individual identifier variable.
  \item moment: moment identifier variable (T0, T1, T2).
  \item group: group identifier variable (placebo, probiotic).
  \item YFAS: proportion of symptoms that characterize addiction.
  \item BES: BES score transformed to unit scale.
\end{itemize}

\begin{figure}[h]
\centerline{\includegraphics[scale = 0.9]{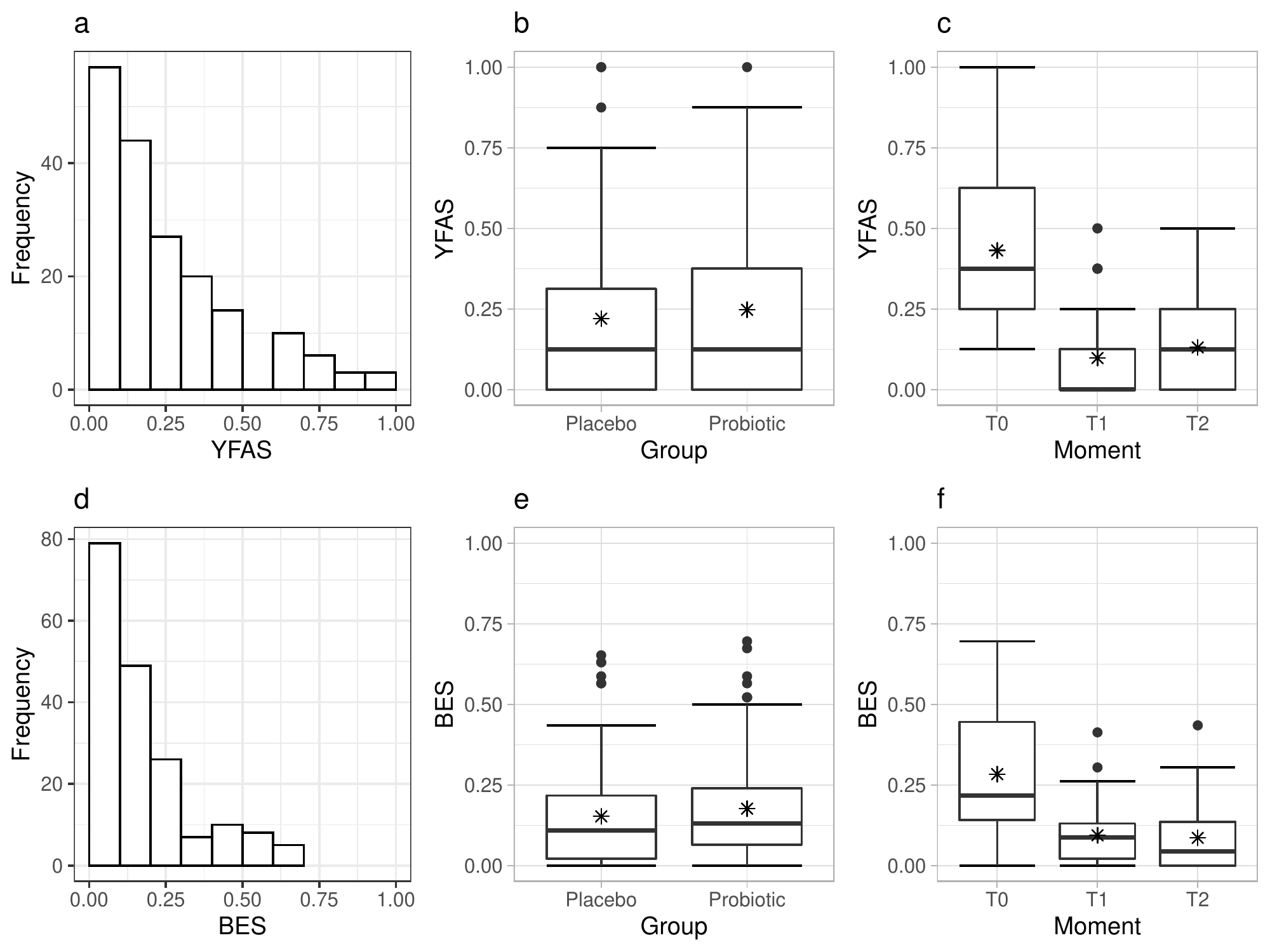}}
\caption{Exploratory graphic analysis: (a) YFAS histogram, (b) YFAS boxplots as a function of group, (c) YFAS boxplots as a function of moment, (d) BES histogram, (b) BES boxplots as a function of group, (c) BES boxplots as a function of moment. The asterisk in the boxplots indicates the mean.\label{fig1}}
\end{figure}

The graphical analysis presented in \autoref{fig1} shows, in (a) and (d) that both variables of interest present considerable asymmetry to the right. The boxplots of metrics as a function of group presented in (b) and (e) show little difference between the placebo and probiotic groups for both responses. The boxplots of the metrics as a function of the evaluation moments, presented in (c) and (f), show that for both metrics the values were higher at T0, with a considerable reduction at T1. When comparing T1 and T2, for YFAS it seems that there is a slight increase in the last evaluation; for BES, T1 and T2 do not seem to differ.  

Still in an exploratory way, we can evaluate the behavior of addiction and compulsion metrics through the evaluation of descriptive measures by moment and by group, presented in \autoref{tab:tab1}. It is possible to verify the reduction of individuals over time, something common in prospective studies. Regarding the measurements, it is noted that both groups (placebo or probiotic) have higher means at T0 than at other times. Therefore, there is a clear reduction in metrics when compared to the preoperative period. When comparing the postoperative moments (T1 and T2) we verified that the YFAS measurements for the placebo and probiotic groups showed, on average, an increase in the metrics at the last evaluation moment. The same could be seen for BES in the placebo group. On the other hand, the BES measures in the probiotic group showed a decrease.

\begin{table}[h]
\centering
\begin{tabular}{ccccc}
\hline
\multirow{2}{*}{Group} & \multirow{2}{*}{Moment}  & \multirow{2}{*}{n} & YFAS                  & BES                   \\ \cline{4-5} 
                       &                          &                    & Mean (standard deviation) & Mean (standard deviation) \\ \hline
Placebo                & T0                       & 33                 & 0.37 (0.26)           & 0.24 (0.20)           \\
Placebo                & T1                       & 32                 & 0.11 (0.15)           & 0.09 (0.10)           \\
Placebo                & T2                       & 22                 & 0.16 (0.15)           & 0.10 (0.12)           \\
Probiotic              & T0                       & 38                 & 0.49 (0.24)           & 0.32 (0.18)           \\
Probiotic              & T1                       & 37                 & 0.09 (0.12)           & 0.10 (0.08)           \\
Probiotic              & T2                       & 22                 & 0.10 (0.14)           & 0.07 (0.09)           \\ \hline
\end{tabular}
\caption{Number of individuals, mean and standard deviation for YFAS and BES for each combination of group and time point.}
\label{tab:tab1}
\end{table}


\section{Multivariate Covariance Generalized Linear Models}\label{sec3}

Consider $\boldsymbol{Y}_{N \times R} = \left \{ \boldsymbol{Y}_1, \dots, \boldsymbol{Y}_R \right \}$ be a matrix of response variables and $\boldsymbol{ M}_{N \times R} = \left \{ \boldsymbol{\mu}_1, \dots, \boldsymbol{\mu}_R \right \}$ be a matrix of expected values. The variance and covariance matrix for each response $r$, $r = 1,..., R$, is denoted by $\Sigma_r$ and has dimension $N \times N$. In addition, consider a correlation matrix $\Sigma_b$, of order $R \times R$, which describes the correlation between the response variables.

The McGLMs \citep{Bonat16} are defined by:

$$
      \begin{aligned}
        \mathrm{E}(\boldsymbol{Y}) &=
          \boldsymbol{M} =
            \{g_1^{-1}(\boldsymbol{X}_1 \boldsymbol{\beta}_1),
            \ldots,
            g_R^{-1}(\boldsymbol{X}_R \boldsymbol{\beta}_R)\}
          \\
        \mathrm{Var}(\boldsymbol{Y}) &=
          \boldsymbol{C} =
            \boldsymbol{\Sigma}_R \overset{G} \otimes
            \boldsymbol{\Sigma}_b,
      \end{aligned}
$$

\noindent where the $g_r()$ are standard link functions; $\boldsymbol{X}_r$ denotes a $N \times k_r$ design matrix; $\boldsymbol{\beta}_r$ denotes a  $k_r \times 1$ vector of regression parameters. $\boldsymbol{\Sigma}_R \overset{G} \otimes \boldsymbol{\Sigma}_b = \mathrm{Bdiag}(\tilde{\boldsymbol{\Sigma}}_1, \ldots, \tilde{\boldsymbol{ \Sigma}}_R) (\boldsymbol{\Sigma}_b \otimes \boldsymbol{I}) \mathrm{Bdiag}(\tilde{\boldsymbol{\Sigma}}_1^\top, \ldots, \tilde{\boldsymbol{\Sigma}}_R^\top)$ denotes the generalized Kronecker product \cite{martinez13}, the matrix $\tilde{\boldsymbol{\Sigma}}_r$ denotes the lower triangular matrix of the Cholesky decomposition of the matrix ${\boldsymbol{\Sigma}}_r$. The operator $\mathrm{Bdiag()}$ denotes the block-diagonal matrix and $\boldsymbol{I}$ is an identity matrix $N \times N$.

For continuous, binary, binomial, proportions or indices, the variance and covariance matrix $\boldsymbol{\Sigma}_r$ is given by:

$$
\Sigma_r =
\mathrm{V}\left(\boldsymbol{\mu}_r; p_r\right)^{1/2}(\boldsymbol{\Omega}\left(\boldsymbol{\tau}_r\right))\mathrm{V}\left(\boldsymbol{\mu}_r; p_r\right)^{1/2}.
$$

In the case of response variables that are counts, the variance and covariance matrix for each response variable is given by:

$$
\Sigma_r = diag(\boldsymbol{\mu}_r)+ \mathrm{V}\left(\boldsymbol{\mu}_r; p_r\right)^{1/2}(\boldsymbol{\Omega}\left(\boldsymbol{\tau}_r\right))\mathrm{V}\left(\boldsymbol{\mu}_r; p_r\right)^{1/2},
$$

\noindent where $\mathrm{V}\left(\boldsymbol{\mu}_r; p_r\right) = diag(\vartheta(\boldsymbol{\mu}_r; p_r))$ denotes a diagonal matrix in which the entries are given by the variance function $\vartheta(\cdot; p_r)$ applied to the elements of the vector $\boldsymbol{\mu}_r$. Different choices of variance functions $\vartheta(\cdot; p_r)$ imply different assumptions about the distribution of the response variable. We will mention 3 options of variance functions: power variance function, Poisson–Tweedie dispersion function and binomial variance function.

The power variance function characterizes the Tweedie family of distributions, it is given by $\vartheta\left(\cdot; p_r\right) = \mu^{p_r}_r$, in which some distributions stand out: Normal ($p $ = 0), Poisson ($p$ = 1), gamma ($p$ = 2) and Inverse Gaussian ($p$ = 3) \cite{Jorgensen87, Jorgensen97}. 

The Poisson–Tweedie dispersion function \cite{Jorgensen15} is an option for responses that characterize counts. The dispersion function given by $\vartheta\left(\cdot; p\right) = \mu + \tau\mu^p$ where $\tau$ is the dispersion parameter. Thus, we have a rich class of models to deal with responses that characterize counts, since many important distributions appear as special cases, such as: Hermite ($p$ = 0), Neyman type A ($p$ = 1), negative binomial ($p$ = 2) and Poisson–inverse Gaussian (p = $3$).

Finally, the binomial variance function, given by $\vartheta\left(\cdot; p_r\right) = \mu^{p_{r1}}_r(1 - \mu_r)^{p_{r2}}$ is indicated when the response variable is binary, bounded to an interval, or when there is a number of successes in a number of trials.

It is possible to notice that the power parameter $p$ appears in all the variance/dispersion functions discussed. This parameter is especially important because it is an index that distinguishes among different probability distributions that are important in the modeling context and, for this reason, can be used as a tool for automatic selection of the probability distribution that best fits to the data.

The dispersion matrix $\boldsymbol{\Omega({\tau})}$ describes the part of the covariance within each response variable that does not depend on the mean structure, ie, the correlation structure between the sample observations. Based on the ideas of Anderson et al \cite{Anderson73} and Pourahmadi \cite{Pourahmadi00}, Bonat and Jørgensen \cite{Bonat16} proposed to model the dispersion matrix through a matrix linear predictor combined with a covariance link function given by:

$$
h\left \{ \boldsymbol{\Omega}(\boldsymbol{\tau}_r) \right \} = \tau_{r0}Z_0 + \ldots + \tau_{rD}Z_D,
$$

\noindent where $h()$ is the covariance link function, $Z_{rd}$ with $d$ = 0,$\ldots$, D are known matrices representing the covariance structure present in each response variable $r$ and $\boldsymbol{\tau_r}$ = $(\tau_{r0}, \ldots, \tau_{rD})$ is a vector $(D + 1) \times 1$ of dispersion parameters.

Some possible covariance link functions are identity, inverse and exponential-matrix. The specification of the covariance link function is discussed by Pinheiro and Bates \cite{Pinheiro96} and it is possible to select combinations of matrices to obtain the most well-known models in the literature for longitudinal data, time series, spatial and spatiotemporal data. Further details are discussed by Demidenko \cite{Demidenko13}.

In this way, the McGLMs configure a general framework for analysis via regression models for Gaussian and non-Gaussian data with multiple responses in which no assumptions are made regarding the independence of the observations. The class is defined by 3 functions (link, variance and covariance) in addition to a linear predictor and a matrix linear predictor for each response under analysis.

\subsection{Estimation and inference}

McGLMs are fitted based on the estimating function approach described in detail by Bonat and Jørgensen \cite{Bonat16} and Jørgensen and Knudsen \cite{jorg04}. This subsection presents an overview of the algorithm and the asymptotic distribution of estimators based on estimating functions.

McGLMs' second-moment assumptions allow us to spplit the parameter vector into two sets: $\boldsymbol{\theta} = (\boldsymbol{\beta}^{\top}, \boldsymbol{\lambda}^{\top})^ {\top}$. So $\boldsymbol{\beta} = (\boldsymbol{\beta}_1^\top, \ldots, \boldsymbol{\beta}_R^\top)^\top$ is a $K \times 1$ vector of regression parameters and $\boldsymbol{\lambda} = (\rho_1, \ldots, \rho_{R(R-1)/2}, p_1, \ldots, p_R, \boldsymbol{\tau}_1^\top , \ldots, \boldsymbol{\tau}_R^\top)^\top$ is a $Q \times 1$ vector of dispersion parameters. Also, $\mathcal{Y} = (\boldsymbol{Y}_1^\top, \ldots, \boldsymbol{Y}_R^\top)^\top$ denotes the stacked vector of order $NR \times 1$ from the matrix of response variables $\boldsymbol{Y}_{N \times R}$ and $\mathcal{M} = (\boldsymbol{\mu}_1^\top, \ldots, \boldsymbol{\mu}_R^ \top)^\top$ denotes the stacked vector of order $NR \times 1$ of the expected value matrix $\boldsymbol{M}_{N \times R}$.

To estimate the regression parameters, the following quasi-score function \cite{Liang86} is used, 
$$
\begin{aligned}
  \psi_{\boldsymbol{\beta}}(\boldsymbol{\beta},
  \boldsymbol{\lambda}) = \boldsymbol{D}^\top
  \boldsymbol{C}^{-1}(\mathcal{Y} - \mathcal{M}),
\end{aligned}
$$

\noindent where $\boldsymbol{D} = \nabla_{\boldsymbol{\beta}} \mathcal{M}$ is a $NR \times K$ matrix, and $\nabla_{\boldsymbol{\beta}}$ denotes the gradient operator. Using the quasi-score function the $K \times K$ sensitivity matrix of $\psi_{\boldsymbol{\beta}}$ is given by

$$
\begin{aligned}
S_{\boldsymbol{\beta}} = E(\nabla_{\boldsymbol{\beta} \psi \boldsymbol{\beta}}) = -\boldsymbol{D}^{\top} \boldsymbol{C}^{-1} \boldsymbol{D},
\end{aligned}
$$

\noindent the $K \times K$ variability matrix of $\psi_{\boldsymbol{\beta}}$ is written as

$$
\begin{aligned}
V_{\boldsymbol{\beta}} = VAR(\psi \boldsymbol{\beta}) = \boldsymbol{D}^{\top} \boldsymbol{C}^{-1} \boldsymbol{D}.
\end{aligned}
$$

For the dispersion parameters, the Pearson estimating function is used, defined in the form

$$
  \begin{aligned}
    \psi_{\boldsymbol{\lambda}_i}(\boldsymbol{\beta},
    \boldsymbol{\lambda}) =
    \mathrm{tr}(W_{\boldsymbol{\lambda}i}
    (\boldsymbol{r}^\top\boldsymbol{r} -
    \boldsymbol{C})),  i = 1,.., Q, 
  \end{aligned}
$$

\noindent where $W_{\boldsymbol{\lambda}i} = -\frac{\partial \boldsymbol{C}^{-1}}{\partial \boldsymbol{\lambda}_i}$ and $\boldsymbol{ r} = (\mathcal{Y} - \mathcal{M})$. The entry $(i,j)$ of the sensitivity matrix $Q \times Q$ of $\psi_{\boldsymbol{\lambda}}$ is given by

$$
  \begin{aligned}
    S_{\boldsymbol{\lambda_{ij}}} = E \left (\frac{\partial }{\partial \boldsymbol{\lambda_{i}}} \psi \boldsymbol{\lambda_{j}}\right) = -tr(W_{\boldsymbol{\lambda_{i}}} CW_{\boldsymbol{\lambda_{J}}} C).
  \end{aligned}
$$

\noindent The entry $(i,j)$ of the variability matrix $Q \times Q$ of $\psi_{\boldsymbol{\lambda}}$ is defined by

$$
  \begin{aligned}
V_{\boldsymbol{\lambda_{ij}}} = Cov\left ( \psi_{\boldsymbol{\lambda_{i}}}, \psi_{\boldsymbol{\lambda_{j}}} \right) = 2tr(W_{\boldsymbol{\lambda_{i}}} CW_{\boldsymbol{\lambda_{J}}} C) + \sum_{l=1}^{NR} k_{l}^{(4)} (W_{\boldsymbol{\lambda_{i}}})_{ll} (W_{\boldsymbol{\lambda_{j}}})_{ll},
  \end{aligned}
$$

\noindent where $k_{l}^{(4)}$ denotes the fourth cumulant of $\mathcal{Y}_{l}$. In the McGLM estimation process, empirical versions are used.

To take into account the covariance between the vectors $\boldsymbol{\beta}$ and $\boldsymbol{\lambda}$, Bonat and Jørgensen \cite{Bonat16} obtained the sensitivity and cross-variability matrices, denoted by $S_{\boldsymbol{ \lambda \beta}}$, $S_{\boldsymbol{\beta \lambda}}$ and $V_{\boldsymbol{\lambda \beta}}$, more details in Bonat and Jørgensen \cite{Bonat16}. The joint sensitivity and variability matrices of $\psi_{\boldsymbol{\beta}}$ and $\psi_{\boldsymbol{\lambda}}$ are denoted by

$$
  \begin{aligned}
    S_{\boldsymbol{\theta}} = \begin{bmatrix}
      S_{\boldsymbol{\beta}} & S_{\boldsymbol{\beta\lambda}} \\ 
      S_{\boldsymbol{\lambda\beta}} & S_{\boldsymbol{\lambda}} 
      \end{bmatrix} \text{e } V_{\boldsymbol{\theta}} = \begin{bmatrix}
      V_{\boldsymbol{\beta}} & V^{\top}_{\boldsymbol{\lambda\beta}} \\ 
      V_{\boldsymbol{\lambda\beta}} & V_{\boldsymbol{\lambda}} 
    \end{bmatrix}.
  \end{aligned}
$$

Let $\boldsymbol{\hat{\theta}} = (\boldsymbol{\hat{\beta}^{\top}}, \boldsymbol{\hat{\lambda}^{\top}})^{\top }$ be the estimator based on estimating functions of $\boldsymbol{\theta}$. Then the asymptotic distribution of $\boldsymbol{\hat{\theta}}$ is

$$
  \begin{aligned}
    \boldsymbol{\hat{\theta}} \sim N(\boldsymbol{\theta}, J_{\boldsymbol{\theta}}^{-1}),
  \end{aligned}
$$

\noindent where $J_{\boldsymbol{\theta}}^{-1}$ is the inverse of the Godambe information matrix, given by $J_{\boldsymbol{\theta}}^{-1} = S_{ \boldsymbol{\theta}}^{-1} V_{\boldsymbol{\theta}} S_{\boldsymbol{\theta}}^{-\top}$, where $S_{\boldsymbol{\theta}} ^{-\top} = (S_{\boldsymbol{\theta}}^{-1})^{\top}.$

To solve the system of equations $\psi_{\boldsymbol{\beta}} = 0$ and $\psi_{\boldsymbol{\lambda}} = 0$, Jørgensen and Knudsen \cite{jorg04} proposed the modified Chaser algorithm, which is defined as

$$
\begin{aligned}
\begin{matrix}
\boldsymbol{\beta}^{(i+1)} = \boldsymbol{\beta}^{(i)}- S_{\boldsymbol{\beta}}^{-1} \psi \boldsymbol{\beta} (\boldsymbol{\beta}^{(i)}, \boldsymbol{\lambda}^{(i)}), \\ 
\boldsymbol{\lambda}^{(i+1)} = \boldsymbol{\lambda}^{(i)} - \alpha S_{\boldsymbol{\lambda}}^{-1} \psi \boldsymbol{\lambda} (\boldsymbol{\beta}^{(i+1)}, \boldsymbol{\lambda}^{(i)}).
\end{matrix}
\end{aligned}
$$


\section{Wald Test for McGLMs}\label{sec4}

This section is dedicated to the presentation of our proposal: the use of the Wald test to evaluate McGLMs parameters. It is worth remembering that in McGLMs there are regression, dispersion, power and correlation parameters. Each set of parameters has a relevant practical interpretation in such a way that through the regression parameters it is possible to identify the relevant explanatory variables, through the dispersion parameters it is possible to evaluate the impact of the correlation between units of the data set, through the power parameters it is possible to identify which probability distribution best fits the data according to the variance function, and through the correlation parameters it is possible to assess the association between responses. In this article, we are interested in methods for evaluation of regression and dispersion parameters.


Let $\boldsymbol{\theta^{*}}$ be a $h \times 1$ vector of parameters with the exception of correlation parameters, i.e. $\boldsymbol{\theta^{*}}$ refers only to regression, dispersion or power parameters. 
The parameter estimates of $\boldsymbol{\theta^{*}}$ are given by $\boldsymbol{\hat\theta^{*}}$. Similarly, let $J^{\boldsymbol{*}-1}$ be the $h \times h$ inverse of the Godambe information matrix excluding the correlation parameters. Let $\boldsymbol{L}$ be a $s \times h$ matrix specifying the hypotheses to be tested and $\boldsymbol{c}$ be a $s \times 1$ vector with values under the null hypothesis, where $s$ denotes the number of constraints; the hypotheses to be tested can be written as:

\begin{equation}
\label{eq:hipoteses_wald}
H_0: \boldsymbol{L}\boldsymbol{\theta^{*}} = \boldsymbol{c} \ vs \ H_1: \boldsymbol{L}\boldsymbol{\theta^{*}} \neq \boldsymbol{c}. 
\end{equation}

\noindent In this way, the generalization of the Wald test statistic to verify the validity of a hypothesis about parameters of a McGLM is given by:

$$
W = (\boldsymbol{L\hat\theta^{*}} - \boldsymbol{c})^T \ (\boldsymbol{L \ J^{\boldsymbol{*}-1} \ L^T})^{-1} \ (\boldsymbol{L\hat\theta^{*}} - \boldsymbol{c}),
$$

\noindent where $W \sim \chi^2_s$, that is, regardless of the number of parameters in the hypotheses, the test statistic $W$ is a single value that asymptotically follows the $\chi^2$ distribution with degrees of freedom given by the number of constraints, that is, the number of rows in the matrix $\boldsymbol{L}$, denoted by $s$.

In general, each column of the matrix $\boldsymbol{L}$ corresponds to one of the $h$ parameters of $\boldsymbol{\theta^{*}}$ and each row to a constraint. Its construction basically consists of filling the matrix with 0, 1 and eventually -1 in such a way that the product $\boldsymbol{L}\boldsymbol{\theta^{*}}$ correctly represents the hypotheses of interest. The correct specification of $\boldsymbol{L}$ allows us testing any parameter individually or even formulating hypotheses for several parameters.

In a practical context, after obtaining the estimates of the model parameters, we may be interested in three types of hypotheses: (i) interest in assessing whether there is evidence that allows us to state that only a single parameter is equal to a postulated value; (ii) interest in assessing whether there is evidence to state that a set of parameters is equal to a postulated vector of values; (iii) interest in knowing if the difference between the effects of two variables is equal to 0, that is, if the effect of the variables on the response is the same.

For the purposes of illustrating the types of hypotheses mentioned, consider the situation in which you want to investigate whether a numeric variable $X_1$ has an effect on two response variables, denoted by $Y_1$ and $Y_2$. For this task, consider that a sample with $N$ observations was collected and for each observation the values of $X_1$, $Y_1$ and $Y_2$ were recorded. Based on the collected data, a bivariate McGLM was fitted, with a linear predictor given by:

\begin{equation}
\label{eq:pred_ex}
g_r(\mu_r) = \beta_{r0} + \beta_{r1} X_1, r=1,2,
\end{equation}

\noindent where the index $r$ denotes the response variable, r = 1,2; $\beta_{r0}$ represents the intercept; $\beta_{r1}$ a regression parameter associated with a variable $X_1$. We assume that each response has only one dispersion parameter $\tau_{r0}$ and that the power parameters have been fixed. Therefore, it is a problem in which there are two response variables and only one explanatory variable. We assume that the observations under study are independent, so $Z_0 = I$. 

In this scenario, questions of interest could be: is there an effect of the variable $X_1$ on only one of the responses? Is it possible that the variable $X_1$ has an effect on both responses at the same time? Is it possible that the effect of the variable is the same for both responses? All these questions can be answered by testing hypotheses on the model parameters and specified using \autoref{eq:hipoteses_wald}. In the next subsections, we present the elements necessary to conduct each test.

\subsection{Example 1: hypothesis for a single parameter}

Let's consider the simplest type of hypothesis to test: a hypothesis about a single parameter. We are interested in to evaluate whether there is an effect of the variable $X_1$ only on the first response. The hypothesis can be written as follows:

\begin{equation}
\label{eq:ex1}
H_0: \beta_{11} = 0 \ vs \ H_1: \beta_{11} \neq 0.
\end{equation}

The same hypothesis can be rewritten in the most convenient notation for applying the Wald test statistic, as in \autoref{eq:hipoteses_wald} where:

\begin{itemize}
  
  \item $\boldsymbol{\theta^{*T}}$ = $\begin{bmatrix} \beta_{10} \  \beta_{11} \ \beta_{20} \ \beta_{21} \ \tau_{11} \ \tau_{21} \end{bmatrix}$.

\item $\boldsymbol{L} = \begin{bmatrix} 0 & 1 & 0 & 0 & 0 & 0  \end{bmatrix}.$
 
\item $\boldsymbol{c}$ = $\begin{bmatrix} 0 \end{bmatrix}.$

\end{itemize}

Note that the vector $\boldsymbol{\theta^{*}}$ has six elements, therefore the matrix $\boldsymbol{L}$ contains six columns (one for each element) and one row, because only a single parameter is being tested. This single line is composed of zeros, except for the column referring to the parameter of interest, which receives 1. It is simple to verify that the product $\boldsymbol{L}\boldsymbol{\theta^{*}}$ represents the hypothesis of interest initially postulated in \autoref{eq:ex1}. Thus, the asymptotic distribution of the test is $\chi^2_1$.

\subsection{Example 2: hypothesis for multiple parameters}\label{sec:ex2}

We are interested in to assess whether there is sufficient evidence to state that there is an effect of the explanatory variable $X_1$ on both responses simultaneously. 
In this case, we have to test 2 parameters: $\beta_{11}$, which associates $X_1$ with the first response and $\beta_{21}$, which associates $X_1$ with the second response. 
We can write the hypothesis as follows:

\begin{equation}
\label{eq:ex2}
H_0: \beta_{r1} = 0 \ vs \ H_1: \beta_{r1} \neq 0,
\end{equation}

\noindent or, equivalently:

$$
H_0: 
\begin{pmatrix}
\beta_{11} \\ 
\beta_{21}
\end{pmatrix} 
= 
\begin{pmatrix}
0 \\ 
0
\end{pmatrix}
\ vs \ 
H_1: 
\begin{pmatrix}
\beta_{11} \\ 
\beta_{21}
\end{pmatrix} 
\neq
\begin{pmatrix}
0 \\ 
0 
\end{pmatrix}.
$$

Hypotheses in the form of \autoref{eq:hipoteses_wald} have the following elements:

\begin{itemize}
  
  \item $\boldsymbol{\theta^{*T}}$ = $\begin{bmatrix} \beta_{10} \  \beta_{11} \ \beta_{20} \ \beta_{21} \ \tau_{11} \ \tau_{21} \end{bmatrix}$.

\item $\boldsymbol{L} = \begin{bmatrix} 0 & 1 & 0 & 0 & 0 & 0 \\
0 & 0 & 0 & 1 & 0 & 0 \end{bmatrix}.$
 
\item $\boldsymbol{c} = \begin{bmatrix} 0 \\ 0 \end{bmatrix}.$ 

\end{itemize}

The vector $\boldsymbol{\theta^{*}}$ remains with six elements and the matrix $\boldsymbol{L}$ with six columns. In this case, we are testing two parameters, so the matrix $\boldsymbol{L}$ has two rows. Again, these lines are composed of zeros, except in the columns referring to the parameter of interest. It is simple to verify that the product $\boldsymbol{L}\boldsymbol{\theta^{*}}$ represents the hypothesis of interest initially postulated in \autoref{eq:ex2}. Thus, the asymptotic distribution of the test is $\chi^2_2$.

\subsection{Example 3: hypothesis of equality of parameters}

In this case, we are interested in to testing whether the effect of the variable $X_1$ is the same for both responses. Thus, we form a hypothesis of equality between the parameters or, in other words, if the difference of the effects is null:

\begin{equation}
\label{eq:ex3}
H_0: \beta_{11} - \beta_{21} = 0 \ vs \ H_1: \beta_{11} - \beta_{21} \neq 0,
\end{equation}

\noindent in the notation of \autoref{eq:hipoteses_wald} the elements of the hypotheses are:

\begin{itemize}
  
  \item $\boldsymbol{\theta^{*T}}$ = $\begin{bmatrix} \beta_{10} \  \beta_{11} \ \beta_{20} \ \beta_{21} \ \tau_{11} \ \tau_{21} \end{bmatrix}$.

\item $\boldsymbol{L} = \begin{bmatrix} 0 & 1 & 0 & -1 & 0 & 0  \end{bmatrix}.$
 
\item $\boldsymbol{c}$ = $\begin{bmatrix} 0 \end{bmatrix}.$

\end{itemize}

As there is only one hypothesis, the matrix $\boldsymbol{L}$ has only one row. For the matrix $\boldsymbol{L}$ to be correctly specified in the case of an equality hypothesis, it is necessary to put 1 in the column referring to one parameter, and -1 in the column referring to the other parameter, in such a way that the product $\boldsymbol {L}\boldsymbol{\theta^{*}}$ represents the hypothesis of interest initially postulated. In this case, the product $\boldsymbol{L}\boldsymbol{\theta^{*}}$ generates exactly the same hypothesis specified in \autoref{eq:ex3} and the asymptotic distribution of the test is $\chi^2_1$.

\subsection{Example 4: hypothesis about regression or dispersion parameters for responses under the same predictor}\label{sec:sec_ex4}

The \autoref{eq:pred_ex} describes a generic bivariate model. It is important to note that in this example both responses are subject to the same predictor. In practice, when it comes to McGLMs, different predictors can be specified for each response variables. Thus, what was exposed in \autoref{sec:ex2} is useful for any case in which there is interest in testing hypotheses about more than one model parameter in the same or in different responses, regardless of the predictors between responses.

However, in cases where the responses have identical linear predictors and the assumptions about the parameters do not change from response to response, an alternative specification of the procedure is to use the Kronecker product as used in Bonat et al \cite{plastica}.

Suppose that, in this example, the hypotheses of interest are still written as in \autoref{eq:ex2}. However, as this is a bivariate model with the same predictor for the two responses and the hypothesis of interest are the same for each response and involves only regression coefficients, it is convenient to write the matrix $\boldsymbol{L}$ as the Kronecker product from two matrices: a matrix $\boldsymbol{G}$ and a matrix $\boldsymbol{F}$, that is, $\boldsymbol{L}$ = $\boldsymbol{G} \otimes \boldsymbol{F}$. Thus, the matrix $\boldsymbol{G}$ has dimension $R \times R$ and specifies the hypotheses regarding the responses, whereas the matrix $\boldsymbol{F}$ specifies the hypotheses between variables and has dimension ${s} ' \times {h}'$, where ${s}'$ is the number of linear constraints and ${h}'$ is the total number of coefficients of regression or dispersion of the each response. Therefore, the matrix $\boldsymbol{L}$ has dimension (${s}'R \times h$).

In general, the matrix $\boldsymbol{G}$ is an identity matrix with dimension equal to the number of responses analyzed in the model. Whereas the matrix $\boldsymbol{F}$ is equivalent to a matrix $\boldsymbol{L}$ considering there was only a single response in the model and only regression or dispersion parameters. We use the Kronecker product of these two matrices to ensure that the hypothesis described in the $\boldsymbol{F}$ matrix will be tested on the $R$ model responses.

Thus, considering that this is the case in which the hypotheses can be rewritten by decomposing the $\boldsymbol{L}$ matrix, the test elements are given by:

\begin{itemize}
  
  \item $\boldsymbol{\beta^{T}}$ = $\begin{bmatrix} \beta_{10} \  \beta_{11} \  \beta_{20} \  \beta_{21} \end{bmatrix}$: the regression parameters of the model.

\item $\boldsymbol{G} = \begin{bmatrix} 1 & 0 \\ 0 & 1  \end{bmatrix}$: identity matrix with dimension given by the number of responses.

\item $\boldsymbol{F} = \begin{bmatrix} 0 & 1 \end{bmatrix}$: equivalent to a $\boldsymbol{L}$ for a single response.

\item $\boldsymbol{L} = \boldsymbol{G} \otimes \boldsymbol{F} =  \begin{bmatrix} 0 & 1 & 0 & 0 \\
0 & 0 & 0 & 1 \end{bmatrix}$: matrix specifying the hypotheses on all responses.
 
\item $\boldsymbol{c} = \begin{bmatrix} 0 \\ 0 \end{bmatrix}$: vector of values under the null hypothesis.

\end{itemize}

Thus, the product $\boldsymbol{L}\boldsymbol{\beta}$ represents the initially postulated hypothesis of interest. In this case, the asymptotic distribution of the test is $\chi^2_2$. The procedure is easily generalized when there is interest in evaluating a hypothesis about the dispersion parameters and this specification is quite convenient for generating analysis of variance tables.

\subsection{ANOVA and MANOVA via Wald test}

Based on the proposal to use the Wald test for McGLMs, we aim to propose in this article three different procedures for generating ANOVA and MANOVA tables for regression parameters, following the nomenclature types I, II and III. In addition, we also aim to propose a procedure similar to ANOVA and MANOVA to evaluate the dispersion parameters of a given model. In the case of ANOVAs, a table is generated for each response variable. For MANOVAs only one tables is generated, therefore, in order to be able to perform MANOVAs, the model responses must be subject to the same linear predictor.

With the purposes of to illustrate the tests performed by each type of analysis of variance, we consider the situation in which one wants to investigate whether two numeric variables denoted by $X_1$ and $X_2$ have an effect on two response variables denoted by $Y_1$ and $Y_2$. For this task, we suppose that a sample with $N$ observations was collected and for each observation the values of $X_1$, $X_2$, $Y_1$ and $Y_2$ were recorded. Based on the collected data, a bivariate model was fitted, with linear predictor given by:

$$
g_r(\mu_r) = \beta_{r0} + \beta_{r1} X_1 + \beta_{r2} x_2 + \beta_{r3} X_1X_2.
$$

\noindent where the index $r$ denotes the response variable, r = 1,2; $\beta_{r0}$ represents the intercept; $\beta_{r1}$ a regression parameter associated with the variable $X_1$, $\beta_{r2}$ a regression parameter associated with the variable $X_2$ and $\beta_{r3}$ a regression parameter associated with the interaction between $X_1$ and $X_2$. Assume that the units under study are independent, so each response has only one dispersion parameter $\tau_{r0}$ associated with a matrix $Z_0 = I$. Finally, we consider that the power parameters have been fixed.

\subsubsection{ANOVA and MANOVA type I}

Our type I analysis of variance proposal for McGLMs performs tests on the regression parameters sequentially. In this scenario, the following tests would be performed:

\begin{enumerate}
   \item Tests if all parameters are equal to 0.
   \item Tests if all parameters except intercept are equal to 0.
   \item Tests if all parameters except intercept and parameters referring to $X_1$ are equal to 0.
   \item Tests if all parameters except intercept and parameters referring to $X_1$ and $X_2$ are equal to 0.
\end{enumerate}

Each of these tests would be a row of the analysis of variance table. In the case of ANOVA, one table per response would be generated, in the case of MANOVA a table in which the hypotheses are tested for both responses. This procedure can be called sequential because a variable is added to each line. In general, precisely because of this sequentiality, it is difficult to interpret the effects of variables using type I analysis of variance. On the other hand, type II and III analyzes test hypotheses that are generally of more interest.

\subsubsection{ANOVA and MANOVA type II}

Our type II analysis of variance performs tests similar to the last test of the sequential analysis of variance. In a model without interactions what is done is, in each line, testing the complete model against the model without a variable. In this way, the effect of that variable on the complete model becomes better interpretable, that is, the impact on the quality of the model if we removed a certain variable.

If there are interactions in the model, the complete model is tested against the model without the main effect and any interaction effect involving the variable. Considering the previous example, the type II analysis of variance would do the following tests:

\begin{enumerate}
  \item Tests if the intercept is equal to 0.
  
  \item Tests if the parameters referring to $X_1$ are equal to 0. That is, the impact of removing $X_1$ from the model is evaluated. In this case, the interaction is removed because it contains $X_1$.
  
  \item Tests if the parameters referring to $X_2$ are equal to 0. That is, the impact of removing $X_2$ from the model is evaluated. In this case, the interaction is removed because it contains $X_2$.
  
  \item Tests if the interaction effect is 0.

\end{enumerate}

Note that, in the lines that we aim to understand the effect of $X_1$ and $X_2$, the interaction is also evaluated, as all parameters involving that variable are removed from the model.

\subsubsection{ANOVA and MANOVA type III}

Finally, our type III analysis of variance considers the complete model against the model without a certain effect. Considering the previous example, the type III analysis of variance would do the following tests:

\begin{enumerate}
  \item Tests if the intercept is equal to 0.
  
  \item Tests if the main effect parameters referring to $X_1$ are equal to 0. That is, the impact of withdrawing $X_1$ on the main effects of the model is evaluated. In this case, unlike type II, nothing is assumed about the interaction parameter, even though it involves $X_1$.
  
  \item Tests if the main effect parameters referring to $X_2$ are equal to 0. That is, the impact of withdrawing $X_2$ on the main effects of the model is evaluated. In this case, unlike type II, nothing is assumed about the interaction parameter, even though it involves $X_2$.
  
  \item Tests if the interaction effect is 0.
\end{enumerate}

Note that in the lines where the effect of $X_1$ and $X_2$ is tested, the interaction effect is maintained, unlike what is done in the type II analysis of variance. It is important to note that the type II and III analyzes of variance as proposed in this work generate the same results when applied to models without interaction effects. Furthermore, the procedures can be easily generalized to deal with dispersion parameters.

\subsection{Multiple comparisons test via Wald test}

When ANOVA shows a significant effect of a categorical variable, it is usually of interest to assess which of the levels differ from each other. For this, multiple comparison tests are used. In the literature there are several procedures to perform such tests, many of them described in Hsu \cite{hsu1996multiple}.

Such a situation can be evaluated using the Wald test. By correctly specifying the $\boldsymbol{L}$ matrix, it is possible to evaluate hypotheses about any possible contrast between the levels of a given categorical variable. Therefore, it is possible to use Wald's statistics to perform multiple comparison tests as well.

The procedure is basically based on 3 steps. The first one is to obtain the matrix of linear combinations of the model parameters that result in the fitted means. With this matrix it is possible to generate the matrix of contrasts, given by subtracting each pair of lines from the matrix of linear combinations. Finally, just select the lines of interest from this matrix and use them as the Wald test hypothesis specification matrix, in place of the $\boldsymbol{L}$ matrix.
	
For example, suppose there is a response variable $Y$ subject to an explanatory variable $X$ of 4 levels: A, B, C and D. Consider, to evaluate the effect of the variable $X$, a model given by:

$$g(\mu) = \beta_0 + \beta_1[X=B] + \beta_2[X=C] + \beta_3[X=D].$$

\noindent In this parameterization, the first level of the categorical variable is used as a reference category and, for the other levels, the change to the reference category is measured; this is called treatment contrast. In this context, $\beta_0$ represents the fitted mean of level A, while $\beta_1$ represents the difference from A to B, $\beta_2$ represents the difference from A to C and $\beta_3$ represents the difference from A to D. With this parameterization it is possible to obtain the predicted value for any of the categories in such a way that if the individual belongs to category A, $\beta_0$ represents the predicted value; if the individual belongs to category B, $\beta_0 + \beta_1$ represents the predicted value; for category C, $\beta_0 + \beta_2$ represents the predicted value, and finally, for category D, $\beta_0 + \beta_3$ represents the predicted value.

In matrix terms, these results can be described as

$$
    \boldsymbol{K_0} = 
      \begin{matrix}
        A\\ 
        B\\ 
        C\\ 
        D 
      \end{matrix} 
    \begin{bmatrix}
      1 & 0 & 0 & 0\\ 
      1 & 1 & 0 & 0\\ 
      1 & 0 & 1 & 0\\ 
      1 & 0 & 0 & 1 
    \end{bmatrix}
$$

Note that the product $\boldsymbol{K_0} \boldsymbol{\beta}$ generates the vector of predictions for each level of $X$. By subtracting the rows from the matrix of linear combinations $\boldsymbol{K_0}$ we generate a matrix of contrasts $\boldsymbol{K_1}$

$$
    \boldsymbol{K_1} = 
      \begin{matrix}
        A-B\\ 
        A-C\\ 
        A-D\\ 
        B-C\\
        B-D\\
        C-D\\ 
      \end{matrix} 
    \begin{bmatrix}
      0 & -1 &  0 &  0\\ 
      0 &  0 & -1 &  0\\ 
      0 &  0 &  0 & -1\\ 
      0 &  1 & -1 &  0\\ 
      0 &  1 &  0 & -1\\ 
      0 &  0 &  1 & -1 
    \end{bmatrix}
$$

To carry out a test of multiple comparisons, it is enough to select the desired contrasts in the lines of the matrix $\boldsymbol{K_1}$ and use these lines as a matrix for specifying the hypotheses of the Wald test. Finally, as usual in tests of multiple comparisons, correction of p-values by means of Bonferroni correction is recommended.

To carry out this procedure for McGLMs, we must remember that this is a class of multivariate models. Thus, as in the case of analysis of variance, for tests of multiple comparisons there are two possibilities: tests for a single response and tests for multiple responses. 

In practice, if the interest is a multivariate multiple comparison test, there is a need for all responses to be subject to the same linear predictor and it is enough to expand the contrast matrix using the Kronecker product, following an idea similar to that exposed in \autoref{sec:sec_ex4}. In the case of a multiple comparison test for each response, simply select the vector of estimates and the partition corresponding to the matrix  $J_{\boldsymbol{\theta}}^{-1}$ for the specific response and proceed with the test. In this way, it is possible to obtain a simple and useful procedure of multiple comparisons when there is a McGLM with categorical explanatory variables and there is an interest in determining which levels differ from each other.


\section{Simulation studies}\label{sec5}

In order to evaluate the power of the Wald test in hypothesis testing on McGLMs parameters, simulation studies were performed. In these simulations, we evaluated the characteristics of the proposal for three probability distributions: Normal, Poisson and Bernoulli. We simulate univariate and trivariate scenarios with different sample sizes to verify the properties of the tests on regression and dispersion parameters. Simulation studies were conducted in the R software \cite{R}. Standard R libraries were used to simulate univariate datasets. To simulate datasets with multiple responses following Normal distribution, the R library \emph{mvtnorm} \cite{mvtnorm1} \cite{mvtnorm2} was used. For the other distributions, the NORTA \cite{norta} method as implemented in the R library \emph{NORTARA} \cite{nortara} was used.

\subsection{Regression parameters}

To evaluate hypotheses concerning regression parameters, sample sizes of 50, 100, 250, 500 and 1000 were considered. 500 data sets were generated for each sample size, simulating a situation with a 4-level categorical explanatory variable. For Normal distribution the regression parameters used were: $\beta_0 = 5$, $\beta_1 = 0$, $\beta_2 = 0$, $\beta_3 = 0$. For the Poisson distribution the regression parameters used were: $\beta_0 = 2.3$, $\beta_1 = 0$, $\beta_2 = 0$, $\beta_3 = 0$. Finally, for the Bernoulli distribution the regression parameters used were: $\beta_0 = 0.5$, $\beta_1 = 0$, $\beta_2 = 0$, $\beta_3 = 0$. The values were chosen in such a way that the coefficient of variation for Normal distribution was 20\%, the Poisson rate were close to 10 and the probability of success for Bernoulli was approximately 0.6. Univariate and trivariate scenarios with these characteristics were evaluated. For trivariate scenarios, there are 4 parameters per response that follow the described settings. For each generated sample, a McGLM was fitted in which the link and variance functions for each distribution are presented in \autoref{tab2}.

\begin{table}[h]
\centering
\begin{tabular}{ccc}
\hline
Distribution & Link function     & Variance function \\ \hline
Normal       & Identity          & Constant           \\
Poisson      & Logarithmic       & Tweedie             \\
Bernoulli    & Logit             & Binomial            \\ \hline
\end{tabular}
\caption{Link and variance functions used in the models for each simulated distribution.}
\label{tab2}
\end{table}

In all cases the matrix predictor for the variance and covariance matrix was specified in order to make explicit that the observations are independent within each response. The correlation between responses in the trivariate case is given by the matrix $\Sigma_b$ described in \autoref{eq:correlacao}.

\begin{equation} \label{eq:correlacao}
\Sigma_b = 
\begin{bmatrix}
1    & 0.75 & 0.5  \\
0.75 & 1    & 0.25 \\
0.5  & 0.25 & 1    \\
\end{bmatrix}
\end{equation}

With the fitted models, the procedure consisted of varying the hypotheses tested on the simulated parameters. We consider 20 different hypotheses based on a decrease $\beta_0$ and even distribute the change along with the other $\beta$s. The decrease for responses following Normal distribution was 0.15; for Poisson distribution the decrease was 0.05; and for Bernoulli distribution the decrease was 0.25. These values were chosen taking into account the desired departure from the tested hypotheses in the response scale and it is important to note that these values are different due to the impact of the link function used in each model.

For each point, we evaluated the percentage of rejection of the null hypothesis. The idea is to verify what happens when we move away from the null hypotheses. It is expected that at the first point there will be a low rejection rate, since the null hypothesis corresponds to the real values of the parameters. For the other points, it is expected that the rate of rejection will increase gradually, as the hypotheses move away and further away from the originally simulated values. The hypotheses tested in each scenario are available in the appendix.
 
\begin{figure}[h]
\centerline{\includegraphics[scale = 0.9]{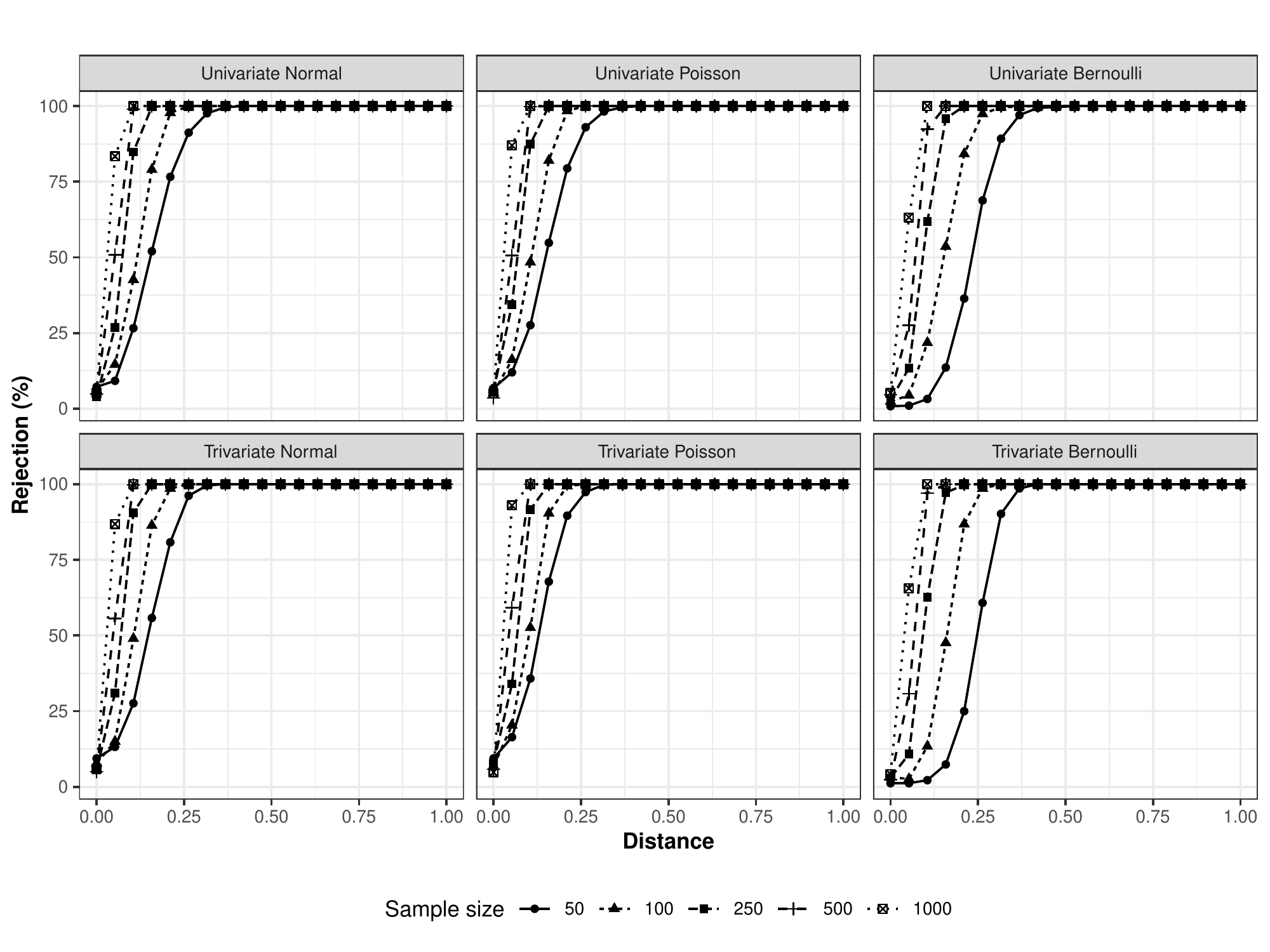}}
\caption{Simulation study results for the regression parameters.\label{fig2}}
\end{figure}

In order to graphically represent the results, we take the Euclidean distance of each hypothesis vector with respect to the vector used to simulate the data. Additionally, we standardized the distance vector by the longest distance to obtain distances between 0 and 1, regardless of the regression parameters. The results are shown in \autoref{fig2}.

In general, the further the hypothesis is from the initially simulated values, the greater the rejection rate. As expected, the lowest rates were observed in the hypothesis equal to the simulated values. In univariate scenarios, the rejection rate was close to 5\% when the hypothesis was equal to the simulated values even with reduced sample sizes. For the trivariate scenarios, in the smallest evaluated sample size, the rejection percentage did not exceed 10\% and in sample sizes equal to 500, the rejection percentage was close to 5\%. Also as expected, it was possible to verify that as the sample size increases, the rejection rate increases for hypotheses that are little different from the simulated values of the parameters.

\subsection{Dispersion parameters}

For the evaluation of hypotheses concerning dispersion parameters, the same sample sizes were considered: 50, 100, 250, 500 and 1000. However, the data sets simulate a situation in which each sample unit provides 5 measurements to the data set. 500 data sets were generated for each sample size and distribution. For Normal distribution, vectors with mean 5 and standard deviation of 1 were simulated. For Poisson distribution, counts with rate equal to 10 were simulated. For Bernoulli distribution, vectors of a dichotomous variable with probability of success equal to 0.6 were simulated.

In all cases, the dispersion parameters to generate the data sets were set at $\tau_0 = 1$, $\tau_1 = 0$ and the effect of explanatory variables was not included. Univariate and trivariate scenarios with these characteristics were evaluated. For each generated sample, a McGLM was fitted with link and variance functions as described in \autoref{tab2}. In trivariate scenarios the correlation between responses is given by \autoref{eq:correlacao}.

In this case, as the objective is to evaluate the correlation within the responses, it is necessary to specify a matrix predictor. The objective is to test hypotheses about the dispersion parameters associated with this matrix predictor. 

With the models adjusted, the procedure consisted of varying the hypotheses tested on the simulated parameters. We consider 20 different hypotheses based on a successive decrease of 0.02 in $\tau_0$ and an increase of 0.02 in $\tau_1$ for each null hypothesis tested. For each point, we evaluated the percentage of rejection of the null hypothesis. The idea is to successively remove the hypothesis from the simulated values and assess whether as we remove the hypothesis from the true values, the percentage of rejection increases. The tested hypotheses are available in the appendix.

\begin{figure}[h]
\centerline{\includegraphics[scale = 0.9]{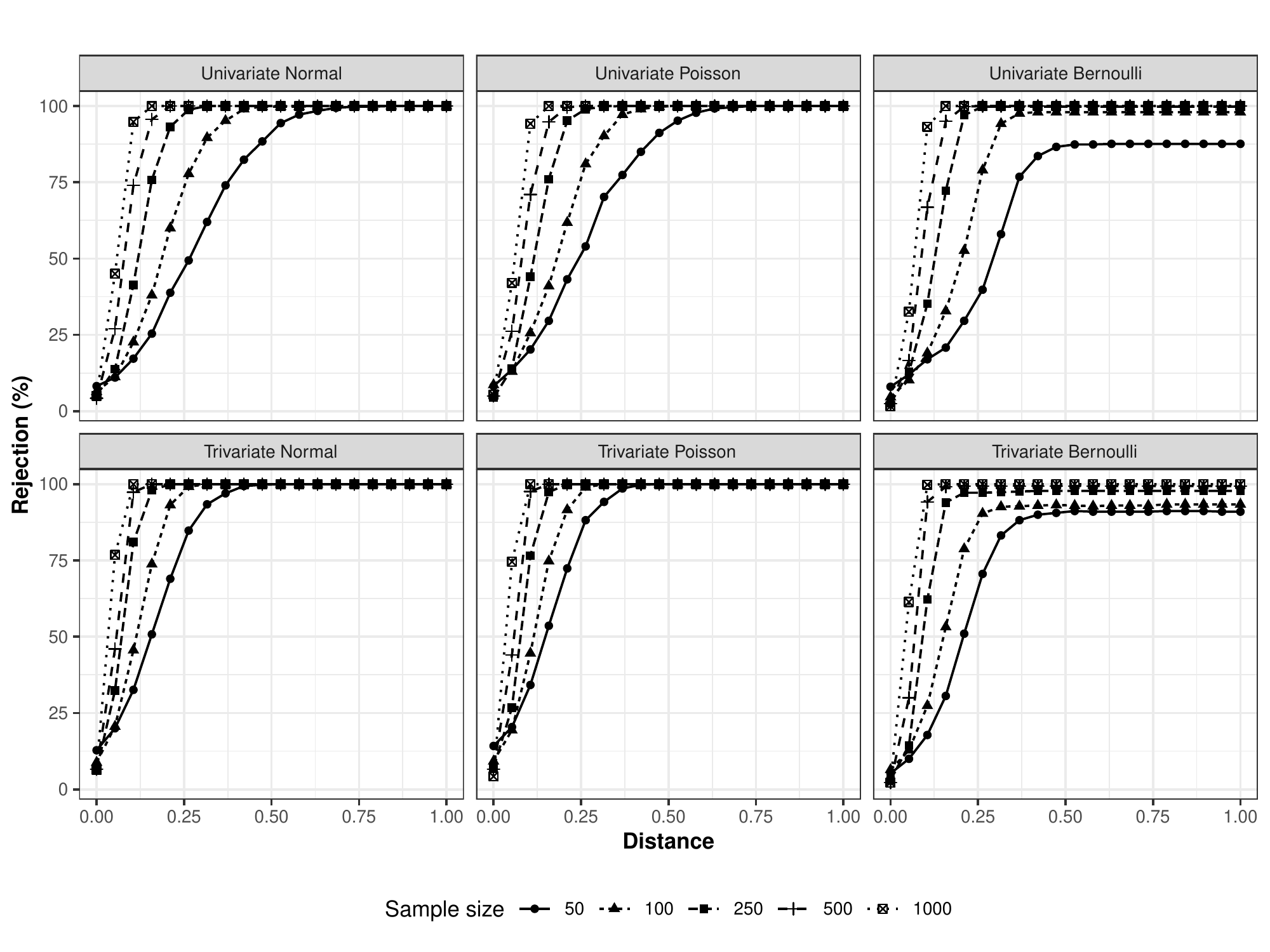}}
\caption{Simulation study results for the dispersion parameters.\label{fig3}}
\end{figure}

In the same way as for the regression parameters, the Euclidean distance of each hypothesis vector in relation to the vector used to simulate the data was taken; and the distance vector was standardized to obtain distances between 0 and 1. The results are shown in \autoref{fig3}.

As observed for the regression parameters, the graphs show that the further the hypothesis is from the initially simulated values, the greater the rejection percentage, and the lower percentages are observed in hypotheses close to the simulated values. In the first hypothesis tested, for the univariate scenarios, rejection percentages close to 8\% were observed in the smallest sample size evaluated. From sample sizes equal to 250, the rejection percentage was close to 5\%. In trivariate cases, the rejection percentage exceeded 10\% in the smallest sample size; for larger sample sizes, the percentages were around 7\%. It was also verified for the dispersion parameters that as the sample size increases, the rejection percentage increases for hypotheses little different from the simulated values of the parameters.


\section{Application}\label{sec6}


In this section we apply the MCGLM approach to analyze the data of the clinical trial presented in \autoref{sec2}, about the effect of use of probiotics in the control of addiction and binge eating disorder in patients undergoing bariatric surgery.

In this application, for composing the linear predictor, we considered the fixed effects of moment (T0,T1,T2) and group (placebo, probiotc). Additionally, the effect of the interaction between these two explanatory variables was included in the model. 

Both responses were treated as proportions, for this reason the logit link function with the binomial variance function was used. Additionally, the power parameter was estimated for both responses under analysis. The linear predictors are given by

$$
g_{r}(\mu_{r}) = \beta_{0r} + \beta_{1r} T1 + \beta_{2r} T2 + \beta_{3r} Probiotic + \\ \beta_{4r} T1*Probiotic + \beta_{5r} T2*Probiotic,
$$

\noindent where the $r$ index refers to the study response variables (1 for YFAS, 2 for BES). The placebo group and the first time point (T0) moment were considered as reference categories. $\beta_{0r}$ represents the intercept, $\beta_{1r}$ the effect of second time point (T1), $\beta_{2r}$ the effect of last time point (T2), $\beta_{3r}$ the probiotic effect. The parameters $\beta_{4r}$ and $\beta_{5r}$ refer to the interaction between moment and group, such that $\beta_{4r}$ represents the effect of the interaction between T1 and probiotic, and $ \beta_{5r}$ represents the effect of the interaction between T2 and probiotic.

As already mentioned, this is an experiment in which the observations are not independent because measurements taken on the same individual are correlated. In McGLMs this correlation can be specified in the model through the matrix linear predictor.

The matrix predictors, the same for both responses, are given by $h\left \{ \boldsymbol{\Omega}(\boldsymbol{\tau}) \right \} = \tau_0Z_0 + \tau_1Z_1$, where the function $h(.)$ used was the identity. The parameters $\tau_0$ and $\tau_1$ represent the dispersion parameters. The matrix $Z_0$ represents an $184 \times 184$ identity matrix and $Z_1$ represents a matrix of dimension $184 \times 184$ specified in order to make it clear that measurements taken from the same individual are correlated. 

To exemplify the form of the matrix predictor, let's consider 3 individuals: A, B and C. Suppose that individual A attended the 3 appointments, so we have information on this individual at T0, T1 and T2. Subject B attended at T0 and T1. The individual C appeared only at T0. So we have 3 individuals and 6 observations. So $Z_0$ is an identity matrix $6 \times 6$ and $Z_1$ is a kind of diagonal block matrix in which the size of the blocks varies according to the number of measurements for each individual. In this scenario, the matrix predictor has the form

\begin{equation}
h\left \{ \boldsymbol{\Omega}(\boldsymbol{\tau}) \right \} = 
\tau_0 \begin{bmatrix}
1 & 0 & 0 & 0 & 0 & 0\\ 
0 & 1 & 0 & 0 & 0 & 0\\ 
0 & 0 & 1 & 0 & 0 & 0\\ 
0 & 0 & 0 & 1 & 0 & 0\\ 
0 & 0 & 0 & 0 & 1 & 0\\ 
0 & 0 & 0 & 0 & 0 & 1\\ 
\end{bmatrix} + 
\tau_1 \begin{bmatrix}
1 & 1 & 1 & 0 & 0 & 0\\ 
1 & 1 & 1 & 0 & 0 & 0\\ 
1 & 1 & 1 & 0 & 0 & 0\\ 
0 & 0 & 0 & 1 & 1 & 0\\ 
0 & 0 & 0 & 1 & 1 & 0\\ 
0 & 0 & 0 & 0 & 0 & 1\\ 
\end{bmatrix}.
\end{equation}

\begin{figure}[h]
\centerline{\includegraphics[scale = 0.9]{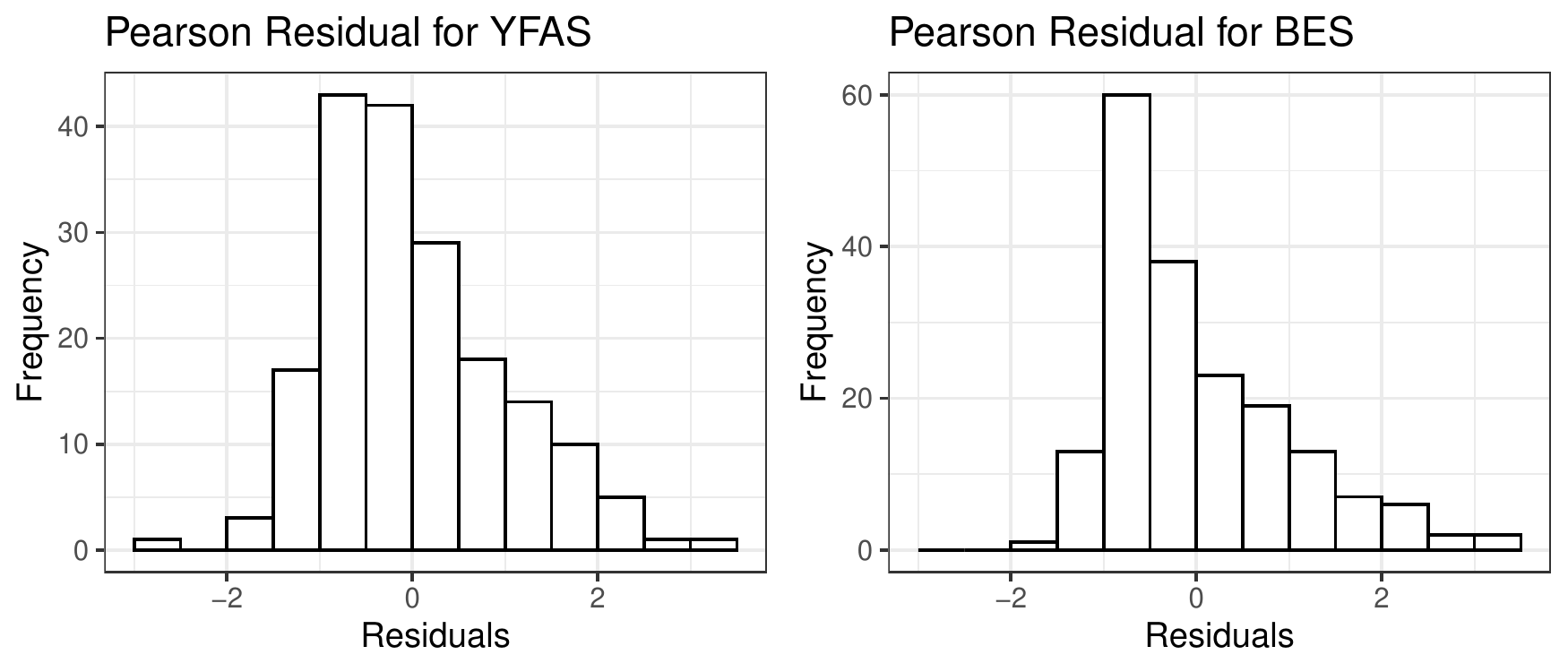}}
\caption{Histogram of Pearson residuals by response.\label{fig4}}
\end{figure}
\begin{figure}[h]
\centerline{\includegraphics[scale = 0.9]{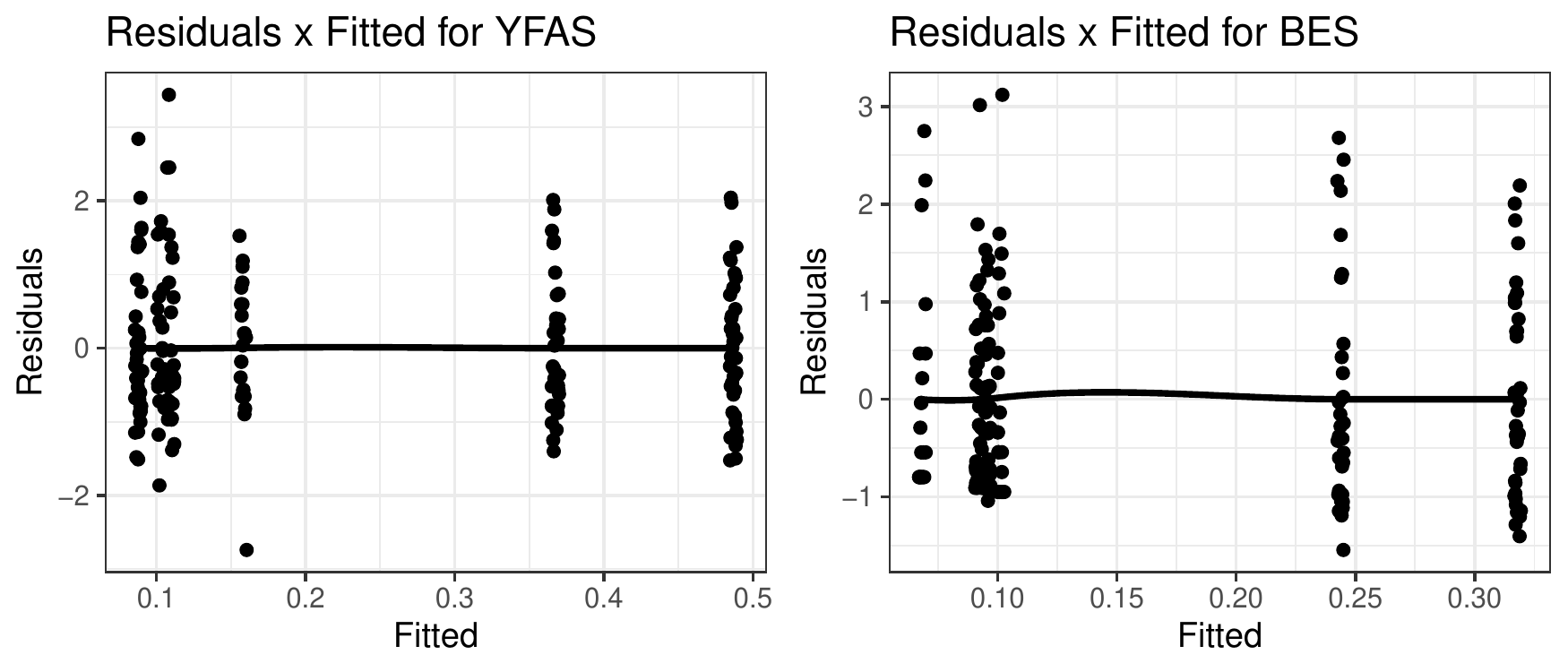}}
\caption{Pearson residuals versus predicted plot with smooth trend line for each response.\label{fig5}}
\end{figure}

We fitted the model and, in order to verify the quality of the model's fit, the residual analysis of the model was performed. The histograms of Pearson's residuals per response, in \autoref{fig4}, show that the distribution of residuals is approximately symmetrical with most of the data between -2 and 2. The predicted versus residuals of the model, presented in \autoref{fig5}, show that there does not seem to be any kind of relationship between residuals and predicted. In addition, the analysis shows that the Pearson residuals for YFAS and BES have a mean of 0 and a standard deviation close to 1. Overall, the model appears to be reasonably well fitted to the data. 

\begin{figure}[h]
\centerline{\includegraphics[width=7in,height=4in]{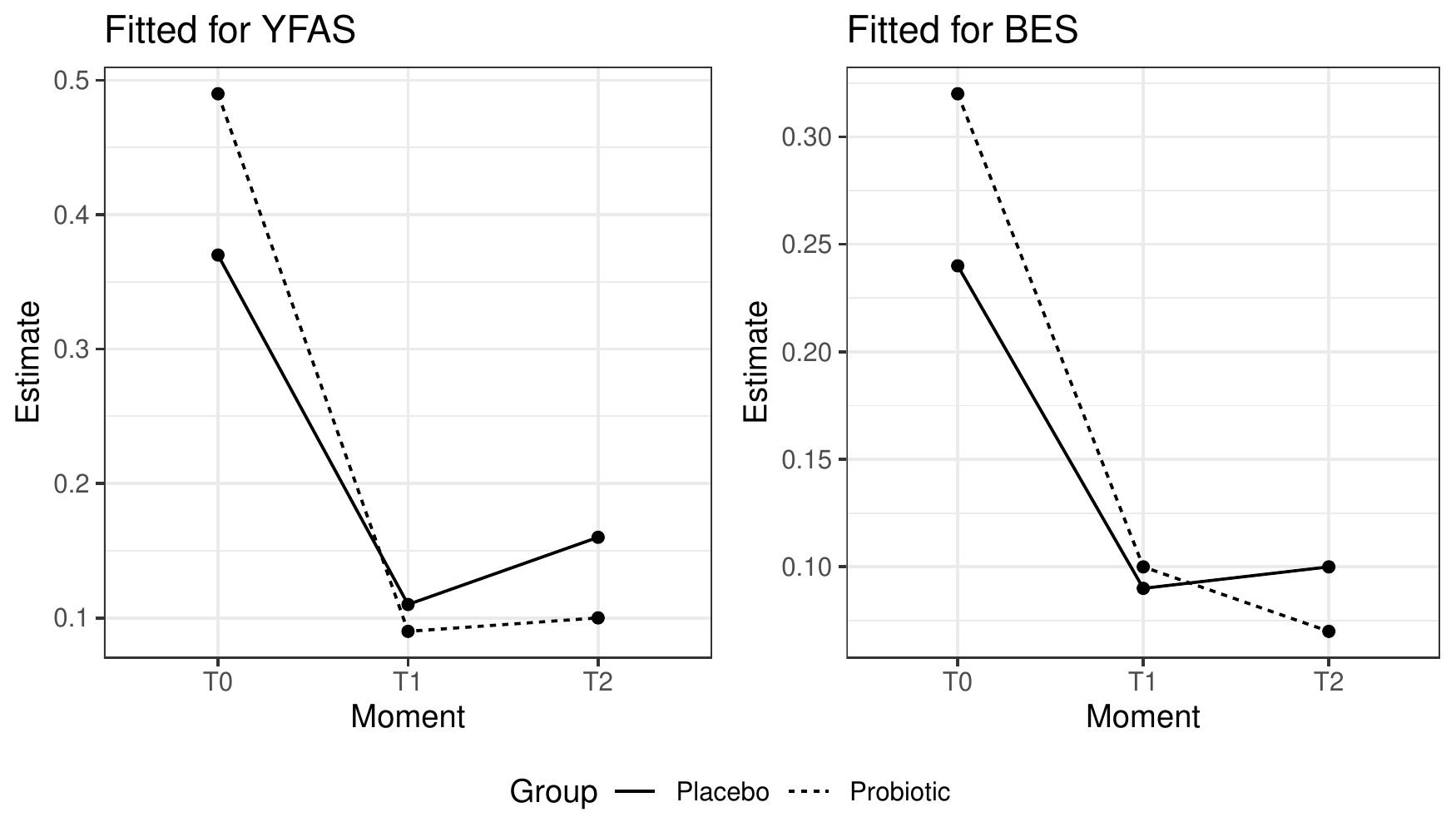}}
\caption{Graph of model predicted for each time and group combination.\label{fig6}}
\end{figure}

In \autoref{tab3} we present the parameter estimates, asymptotic confidence intervals with 95\% confidence, and p-values for the hypothesis that each parameter is equal to zero. Additionally, \autoref{fig6} shows the predicted values for each combination of factors for a better interpretation of the results.

\begin{table}[h]
\centering
\begin{tabular}{c|cccccc}
\hline
\multirow{2}{*}{Parameter} & \multicolumn{3}{c}{YFAS}                                                                                             & \multicolumn{3}{c}{BES}                                                                         \\ \cline{2-7} 
                           & Estimate & \begin{tabular}[c]{@{}c@{}}Confidence \\ interval\end{tabular} & \multicolumn{1}{c|}{p-value}        & Estimate & \begin{tabular}[c]{@{}c@{}}Confidence \\ interval\end{tabular} & p-value        \\ \hline
$\beta_0$                  & -0.54      & (-0.87;-0.22)                                                     & \multicolumn{1}{c|}{\textless 0.01} & -1.13      & (-1.44;-0.83)                                                     & \textless 0.01 \\
$\beta_1$                  & -1.55      & (-2.17;-0.94)                                                     & \multicolumn{1}{c|}{\textless 0.01} & -1.16      & (-1.62;-0.69)                                                     & \textless 0.01 \\
$\beta_2$                  & -1.13      & (-1.75;-0.51)                                                     & \multicolumn{1}{c|}{\textless 0.01} & -1.05      & (-1.58;-0.52)                                                     & \textless 0.01 \\
$\beta_3$                  & 0.49       & (0.05;0.93)                                                       & \multicolumn{1}{c|}{0.0284} & 0.37       & (-0.03;0.77)                                                      & 0.0733           \\
$\beta_4$                  & -0.73      & (-1.60;0.14)                                                      & \multicolumn{1}{c|}{0.0}            & -0.33      & (-0.96;0.30)                                                      & 0.3081           \\
$\beta_5$                  & -0.98      & (-1.93;-0.03)                                                     & \multicolumn{1}{c|}{0.0429} & -0.80      & (-1.58;-0.02)                                                     & 0.0449 \\
$\tau_0$                   & 0.18       & (0.01;0.35)                                                       & \multicolumn{1}{c|}{0.0411} & 0.17       & (0.00;0.34)                                                       & 0.0458           \\
$\tau1$                    & 0.01       & (-0.02;0.04)                                                      & \multicolumn{1}{c|}{0.5718}           & 0.04       & (-0.01;0.10)                                                      & 0.1357           \\
$p$                        & 0.91       & (0.47;1.34)                                                       & \multicolumn{1}{c|}{\textless 0.05} & 1.23       & (0.77;1.68)                                                       & \textless 0.05 \\ \hline
\end{tabular}
\caption{Parameter estimates, 95\% confidence intervals and p-values of the model.}
\label{tab3}
\end{table}

Up to this point, a standard analysis has been presented, with the usual results of the analysis of a regression model. Going further in this analysis, we can make use of the Wald test for a better interpretation of the model parameter estimates.

We can opt for a type II multivariate analysis of variance to assess the importance of the variables in the problem. This result, presented in the table \autoref{tab4}, points to the significant existence of the moment effect and the absence of the group effect, indicating that for both responses the metrics change over time but without change between groups.

\begin{table}[h]
\centering
\begin{tabular}{cccc}
\hline
Variable      & Degrees of freedom & Chi-square   & p-value        \\ \hline
Intercept     & 2                  & 53.1581      & \textless 0.01 \\
Moment        & 8                  & 139.0161     & \textless 0.01 \\
Group         & 6                  & 8.4928       & 0.2042         \\
Moment*Group  & 4                  & 6.9923       & 0.1363         \\ \hline
\end{tabular}
\caption{Type II multivariate analysis of variance.}
\label{tab4}
\end{table}

In order to evaluate the results by response, we can use a type II univariate analysis of variance as shown in \autoref{tab5}. Considering a significance level of $0.01$, there is evidence that points to effect of moment.

\begin{table}[h]
\centering
\begin{tabular}{c|c|cc|cc}
\hline
              &                    & \multicolumn{2}{c|}{YFAS}     & \multicolumn{2}{c}{BES}       \\ \hline
Variable      & Degrees of freedom & Chi-square   & p-value        & Chi-square   & p-value        \\ \hline
Intercept     & 1                  & 10.6128      & \textless 0.01 & 53.1473      & \textless 0.01 \\
Moment        & 4                  & 102.9875     & \textless 0.01 & 99.5681      & \textless 0.01 \\
Group         & 3                  & 6.6837       & 0.0827         & 5.3083       & 0.1506         \\
Moment*Group  & 2                  & 5.5984       & 0.0609         & 4.2477       & 0.1196         \\ \hline
\end{tabular}
\caption{Type II univariate analysis of variance.}
\label{tab5}
\end{table}

As the analysis of variances pointed to a significant effect of categorical variables, we can explore which levels differ from each other. \autoref{tab6} shows two-by-two comparisons between moments. The results show that, for both responses, there are differences between the first versus second and first versus third moments, but the last two moments do not differ from each other. The table \autoref{tab7} presents the comparisons between groups for each time point for both responses. The results point to the absence of difference between groups at each moment.

\begin{table}[h]
\centering
\begin{tabular}{cccc}
\hline
Contrast  & Degrees of freedom & Chi-square   & p-value        \\ \hline
T0-T1     & 2                  & 97.9874      & \textless 0.01 \\
T0-T2     & 2                  & 67.2462      & \textless 0.01 \\
T1-T2     & 2                  & 2.4730       & 0.8712         \\ \hline
\end{tabular}
\caption{Two-by-two comparisons between moments for both responses.}
\label{tab6}
\end{table}

\begin{table}[h]
\centering
\begin{tabular}{cccc}
\hline
Contrast                 & Degrees of freedom & Chi-square   & p-value \\ \hline
T0:Placebo-T0:Probiotic  & 2                  & 5.5819       & 0.9204  \\
T1:Placebo-T1:Probiotic  & 2                  & 0.6096       & 1       \\
T2:Placebo-T2:Probiotic  & 2                  & 1.7645       & 1       \\ \hline
\end{tabular}
\caption{Two-by-two comparisons between groups for each time point for both responses.}
\label{tab7}
\end{table}

In the model, we include the information that there are measurements that were taken on the same individual. This information is declared in the matrix predictor that estimates a dispersion parameter associated with the matrix that indicates the relationship between individuals. A hypothesis of interest may be to assess whether there is evidence to believe that, in this problem, the measurements taken on the same individual are in fact correlated. For this, we can postulate hypotheses about the dispersion parameters. As with analyzes of variance, this can be done per response or for both responses simultaneously. The table \autoref{tab8} presents the results of the multivariate test, that is, it evaluates the hypothesis that in both responses the measures are correlated. The results indicate that there is no evidence to believe that the measures taken in the same individual are correlated.

\begin{table}[h]
\centering
\begin{tabular}{cccc}
\hline
Variable               & Degrees of freedom & Chi-square & p-value        \\ \hline
$\tau_0$ & 2                  & 7.1936       & 0.0274  \\
$\tau_1$ & 2                  & 2.3201       & 0.3135         \\ \hline
\end{tabular}
\caption{Type III multivariate analysis of variance for dispersion parameters.}
\label{tab8}
\end{table}


\section{Concluding remarks}\label{sec7}

The main goal of this article was to develop procedures to perform hypothesis tests on McGLMs parameters based on the Wald statistics. McGLMs rely on regression, dispersion, power, and correlation parameters; each set of parameters has a relevant practical interpretation in the context of data analysis with potential multiple responses as a function of a set of explanatory variables.

We developed procedures for testing general linear hypotheses, generating ANOVA and MANOVA tables for regression and dispersion parameters and also tests of multiple comparisons.

The test properties were evaluated based on simulation studies. Univariate and trivariate scenarios with different probability distributions and sample size were considered.
In order to compute the rejection rate, we simulated 500 dataset under the null hipothesis and gradually move away from it. 
In general, it was possible to observe that the further the hypothesis is from the initially simulated values, the greater the rejection rate. 
As expected, the lowest rates were observed in the hypothesis equal to the simulated values and it was also possible to verify that as the sample size increases, 
the rejection rate increases. Thus, the simulation results showed that the Wald test can be used to evaluate hypotheses about regression parameters and dispersion of McGLMs, 
which allows a better interpretation of the effect of variables and design structures in practical contexts.

Additionally, we applied the proposed methodologies to a dataset in which the objective is to evaluate the effect of the use of probiotics in the control of addiction and binge eating 
disorder in patients undergoing bariatric surgery. This is a clinical trial with two response variables: a score that characterizes compulsion and the number of symptoms presented that characterize addiction.

In this study, a set of subjects was divided into two groups: one of them received a placebo and the other received the treatment. 
In addition, subjects were evaluated over time; in this way, the design generates observations that are not independent, since measurements taken on the same individual tend to be correlated.

The results, based on the tests proposed in this article, indicate that there is evidence that points to a moment effect, that is, addiction and binge eating disorder change over time. 
Multiple comparison tests indicate that, for both responses, there are differences between the first versus second and first versus third moments, but the last two moments do not differ from each other. 
The results also point to the absence of difference between groups at each moment. An evaluation of the dispersion parameters shows that there is no evidence to believe that the measurements taken in the same individual are correlated.

Some limitations of this article concern cases not explored in the simulation studies, such as: evaluation of test performance when defining linear hypotheses that combine parameters of different types, impact of a different number of observations by individuals in longitudinal or repeated measures scenarios, impact on test power as the number of parameters tested increases and evaluate the test quality in multivariate problems with probability distributions different from those explored.

Possible extensions of this article that follow the line of evaluation of McGLMs parameters for a better understanding of the impact of elements in modeling problems are: exploring corrections of p-values according to the size of the tested hypotheses, exploring procedures beyond the Wald test (such as the Score test and the likelihood ratio test), implement new procedures for multiple comparisons, adapt the proposal to deal with alternative contrasts to the usual ones, explore procedures for automatic selection of covariates (backward elimination, forward selection, stepwise selection) and also selection of covariates through the inclusion of penalty in the adjustment for complexity (similar to the spline regression idea). In addition, we intend to develop a package in R language to provide a set of implementations of the procedures presented in this article.



\section*{Acknowledgments}

The authors thank the reviewers for their constructive and helpful comments, which greatly improved the article. This study was financed in part by the Coordenação de Aperfeiçoamento de Pessoal de Nível Superior – Brasil (CAPES) – Finance Code 001.


\subsection*{Conflict of interest}

The authors declare no potential conflict of interests.






\bibliography{draft-sm-lineu.bib}%

\clearpage

\appendix

\section{Hypotheses tested in the simulation study\label{app1}}


\begin{table}[h]
\centering
\begin{tabular}{c|cccc}
\hline
Null hypothesis & $\beta_0$ & $\beta_1$ & $\beta_2$ & $\beta_3$ \\ \hline
$H_{01}$      & 5         & 0         & 0         & 0         \\
$H_{02}$      & 4.85      & 0.05      & 0.05      & 0.05      \\
$H_{03}$      & 4.7       & 0.1       & 0.1       & 0.1       \\
$H_{04}$      & 4.55      & 0.15      & 0.15      & 0.15      \\
$H_{05}$      & 4.4       & 0.2       & 0.2       & 0.2       \\
$H_{06}$      & 4.25      & 0.25      & 0.25      & 0.25      \\
$H_{07}$      & 4.1       & 0.3       & 0.3       & 0.3       \\
$H_{08}$      & 3.95      & 0.35      & 0.35      & 0.35      \\
$H_{09}$      & 3.8       & 0.4       & 0.4       & 0.4       \\
$H_{10}$      & 3.65      & 0.45      & 0.45      & 0.45      \\
$H_{11}$      & 3.5       & 0.5       & 0.5       & 0.5       \\
$H_{12}$      & 3.35      & 0.55      & 0.55      & 0.55      \\
$H_{13}$      & 3.2       & 0.6       & 0.6       & 0.6       \\
$H_{14}$      & 3.05      & 0.65      & 0.65      & 0.65      \\
$H_{15}$      & 2.9       & 0.7       & 0.7       & 0.7       \\
$H_{16}$      & 2.75      & 0.75      & 0.75      & 0.75      \\
$H_{17}$      & 2.6       & 0.8       & 0.8       & 0.8       \\
$H_{18}$      & 2.45      & 0.85      & 0.85      & 0.85      \\
$H_{19}$      & 2.3       & 0.9       & 0.9       & 0.9       \\
$H_{20}$      & 2.15      & 0.95      & 0.95      & 0.95      \\ \hline
\end{tabular}
\caption{Tested hypotheses for regression parameters in models with response following Normal distribution.}
\label{tab:hipoteses_beta_normal}
\end{table}

\begin{table}[h]
\centering
\begin{tabular}{c|cccc}
\hline
Null hypothesis & $\beta_0$ & $\beta_1$ & $\beta_2$ & $\beta_3$ \\ \hline
$H_{01}$      & 2.3       & 0         & 0         & 0         \\
$H_{02}$      & 2.25      & 0.017     & 0.017     & 0.017     \\
$H_{03}$      & 2.2       & 0.033     & 0.033     & 0.033     \\
$H_{04}$      & 2.15      & 0.05      & 0.05      & 0.05      \\
$H_{05}$      & 2.10      & 0.067     & 0.067     & 0.067     \\
$H_{06}$      & 2.05      & 0.083     & 0.083     & 0.083     \\
$H_{07}$      & 2         & 0.1       & 0.1       & 0.1       \\
$H_{08}$      & 1.95      & 0.117     & 0.117     & 0.117     \\
$H_{09}$      & 1.9       & 0.133     & 0.133     & 0.133     \\
$H_{10}$      & 1.85      & 0.15      & 0.15      & 0.15      \\
$H_{11}$      & 1.8       & 0.167     & 0.167     & 0.167     \\
$H_{12}$      & 1.75      & 0.167     & 0.167     & 0.167     \\
$H_{13}$      & 1.7       & 0.2       & 0.2       & 0.2       \\
$H_{14}$      & 1.65      & 0.217     & 0.217     & 0.217     \\
$H_{15}$      & 1.6       & 0.233     & 0.233     & 0.233     \\
$H_{16}$      & 1.55      & 0.25      & 0.25      & 0.25      \\
$H_{17}$      & 1.5       & 0.267     & 0.267     & 0.267     \\
$H_{18}$      & 1.45      & 0.283     & 0.283     & 0.283     \\
$H_{19}$      & 1.4       & 0.3       & 0.3       & 0.3       \\
$H_{20}$      & 1.35      & 0.317     & 0.317     & 0.317      \\ \hline
\end{tabular}
\caption{Tested hypotheses for regression parameters in models with response following Poisson distribution.}
\label{tab:hipoteses_beta_poisson}
\end{table}

\begin{table}[h]
\centering
\begin{tabular}{c|cccc}
\hline
Null hypothesis & $\beta_0$ & $\beta_1$ & $\beta_2$ & $\beta_3$ \\ \hline
$H_{01}$      & 0.5       & 0         & 0         & 0         \\
$H_{02}$      & 0.250     & 0.083     & 0.083     & 0.083     \\
$H_{03}$      & 0         & 0.167     & 0.167     & 0.167     \\
$H_{04}$      & -0.25     & 0.25      & 0.25      & 0.25      \\
$H_{05}$      & -0.500    & 0.333     & 0.333     & 0.333     \\
$H_{06}$      & -0.750    & 0.417     & 0.417     & 0.417     \\
$H_{07}$      & -1.0      & 0.5       & 0.5       & 0.5       \\
$H_{08}$      & -1.250    & 0.583     & 0.583     & 0.583     \\
$H_{09}$      & -1.500    & 0.667     & 0.667     & 0.667     \\
$H_{10}$      & -1.75     & 0.75      & 0.75      & 0.75      \\
$H_{11}$      & -2.000    & 0.833     & 0.833     & 0.833     \\
$H_{12}$      & -2.250    & 0.917     & 0.917     & 0.917     \\
$H_{13}$      & -2.5      & 1.0       & 1.0       & 1.0       \\
$H_{14}$      & -2.750    & 1.083     & 1.083     & 1.083     \\
$H_{15}$      & -3.000    & 1.167     & 1.167     & 1.167     \\
$H_{16}$      & -3.25     & 1.25      & 1.25      & 1.25      \\
$H_{17}$      & -3.500    & 1.333     & 1.333     & 1.333     \\
$H_{18}$      & -3.750    & 1.417     & 1.417     & 1.417     \\
$H_{19}$      & -4.0      & 1.5       & 1.5       & 1.5       \\
$H_{20}$      & -4.250    & 1.583     & 1.583     & 1.583     \\ \hline
\end{tabular}
\caption{Tested hypotheses for regression parameters in models with response following the Bernoulli distribution.}
\label{tab:hipoteses_beta_bernoulli}
\end{table}

\begin{table}[h]
\centering
\begin{tabular}{c|cc}
\hline
Null hypothesis & $\beta_0$ & $\beta_1$ \\ \hline
$H_{01}$      & 1         & 0         \\
$H_{02}$      & 0.98      & 0.02      \\
$H_{03}$      & 0.96      & 0.04      \\
$H_{04}$      & 0.94      & 0.06      \\
$H_{05}$      & 0.92      & 0.08      \\
$H_{06}$      & 0.9       & 0.1       \\
$H_{07}$      & 0.88      & 0.12      \\
$H_{08}$      & 0.86      & 0.14      \\
$H_{09}$      & 0.84      & 0.16      \\
$H_{10}$      & 0.82      & 0.18      \\
$H_{11}$      & 0.8       & 0.2       \\
$H_{12}$      & 0.78      & 0.22      \\
$H_{13}$      & 0.76      & 0.24      \\
$H_{14}$      & 0.74      & 0.26      \\
$H_{15}$      & 0.72      & 0.28      \\
$H_{16}$      & 0.7       & 0.3       \\
$H_{17}$      & 0.68      & 0.32      \\
$H_{18}$      & 0.66      & 0.34      \\
$H_{19}$      & 0.64      & 0.36      \\
$H_{20}$      & 0.62      & 0.38      \\ \hline
\end{tabular}
\caption{Tested hypotheses for dispersion parameters.}
\label{tab:hipoteses_taus}
\end{table}


\end{document}